\documentclass{article}
\usepackage[utf8]{inputenc}
\usepackage[margin=1.2in]{geometry}
\usepackage{graphicx}
\usepackage{authblk}
\usepackage{url}
\usepackage{placeins}
\usepackage{bm}
\usepackage{amsmath}
\usepackage{mathtools}
\usepackage{amsfonts}
\usepackage{xr}
\usepackage{xcolor}

\def\<{\langle}
\def\>{\rangle}
\def\x{{\bf x}}
\def\P{{\cal P}}

\title{Hamiltonian Modeling of Macro-Economic Urban Dynamics}

\author[a]{Bernardo Monechi}
\author[b]{Miguel Ib{\'a}{\~n}ez-Berganza} 
\author[a,b,c]{Vittorio Loreto}

\affil[a]{Sony Computer Science Laboratories, 6, Rue Amyot, 75005, Paris, France}
\affil[b]{Sapienza University of Rome, Physics Department, Piazzale Aldo Moro 2, 00185, Rome, Italy}
\affil[c]{Complexity Science Hub Vienna, Josefst\"adter Strasse 39, A-1080 Vienna, Austria}

\date{}

\begin{document}
\maketitle
\begin{abstract}
	The ongoing rapid urbanization phenomena make the understanding of the evolution of urban environments of utmost importance to improve the well-being and steer societies towards better futures. Many studies have focused on the emerging properties of cities, leading to the discovery of scaling laws mirroring, for instance, the dependence of socio-economic indicators on city sizes. Though scaling laws allow for the definition of city-size independent socio-economic indicators, only a few efforts have been devoted to the modeling of the dynamical evolution of cities as mirrored through socio-economic variables and their mutual influence. In this work, we propose a Maximum Entropy (ME), non-linear, generative model of cities. We write in particular a Hamiltonian function in terms of a few macro-economic variables, whose coupling parameters we infer from real data corresponding to French towns. We first discover that non-linear dependencies among different indicators are needed for a complete statistical description of the non-Gaussian correlations among them. Furthermore, though the dynamics of individual cities are far from being stationary, we show that the coupling parameters corresponding to different years turn out to be quite robust. The quasi time-invariance of the Hamiltonian model allows proposing an analytic model for the evolution in time of the macro-economic variables, based on the Langevin equation. 	Despite no temporal information about the evolution of cities has been used to derive this model, its forecast accuracy of the temporal evolution of the system is compatible to that of a model inferred using explicitly such information. 
\end{abstract}
%
%
%
\section*{Introduction}
One of the significant challenges humanity is currently facing is accelerated urbanisation. According to the UN, some 55 per cent of the global population lives in cities, and this fraction is expected to rise to more than two thirds by 2050. Different scientific communities accepted the challenge and have started to build a deep understanding of the phenomena related to the urban environment, to develop more sustainable and livable cities.
One of the more interesting recent findings in the field of the Science of Cities is the so-called scaling laws in urban indicators. According to these laws, the population $P$ is the crucial determinant for cities, and other macro-economic features of a city, say $X$, depend on $P$ through a power-law $X\sim P^\beta$ with a feature-dependent exponent $\beta$~\cite{bettencourt2013origins, bettencourt2013hypothesis,bettencourt2010urban,youn2016scaling, batty2008size}. Some quantities appear to scale superlinearly with $P$ (i.e., $\beta>1$), for instance, the GDP or the number of serious crimes, while others depend sublinearly on $P$ (i.e., $\beta<1$), e.g., the number of infrastructures~\cite{arcaute2015constructing}. These scaling laws appear as a fundamental property of urban environments, naturally emerging from their growth dynamics~\cite{li2017simple}. Recently, some criticisms have been raised about the concept of scaling in urban systems~\cite{leitao2016scaling}. On the one hand, it has been shown how it is hard to distinguish $X\sim P^\beta$ from a linear dependency on $P$; on the other hand, the exponent $\beta$ might depend on how one defines city boundaries~\cite{cottineau2017diverse,louf2014scaling,oliveira2014large}.

Despite these issues, scaling laws have profound consequences in the way we think about cities. Albeit cities of different size exhibit very different macro-economic features, when described in terms of rescaled variables, they behave in a size-independent way. Consequently, scaling laws allow for a characterisation of cities as abstract, size-independent, entities operating at different scales defined by the population size. Such an intriguing idea has been backed up in time by empirical observations as well as various modelling schemes, trying to grasp the microscopic mechanism responsible for the emergence of scaling. While the identification of the mechanisms behind the emergence of scaling laws is essential to understand the evolution of cities, the current research is still lacking studies aimed at understanding how different indicators influence each other.

To fill this gap, we present here a Maximum Entropy (ME)  generative model for cities written in terms of a few macro-economic variables, whose parameters (the effective Hamiltonian, in a statistical-physical analogy) are inferred from real data through a Maximum Likelihood approach.

In our approach, we assume that scaling laws are an intrinsic property of cities. Focusing on indicators related to the job market (e.g., employment rate, number of jobs in the tertiary, etc.), we exploit scaling-laws to define population-independent macro-economic indicators, through which we construct the model (we refer to the Appendix for further details). The ME inference principle~\cite{mehta2019high} on which our modelling scheme relies,  has a longstanding history of successful applications in statistical physics~\cite{presse2013principles}, biology~\cite{mora2010maximum,bialek2012statistical,mora2016local,lezon2006using}, along with other inter-disciplinary applications ~\cite{stephens2010statistical,sakellariou2017maximum}.  The ME principle guarantees, on rigorous information-theoretical grounds, that the generative model is the most general probability distribution in terms of the considered set of indicators, that reproduces only the statistically significant database statistics, under absence of any other assumption or artifact. The model parameters, i.e., the coupling parameters of the Hamiltonian function, are {\it inferred} following a Maximum Likelihood principle, from a longitudinal dataset composed by about 11000 French ``communes" (the smallest administrative French units ranging from areas of few inhabitants to large metropolis) in $10$ different years.

Our new modelling schemes allows to establish three main results. First, thanks to its non-linear character, the inferred generative model goes beyond the multivariate Gaussian distribution of the indicators and allows us to reproduce the non-trivial empirical correlations among rescaled features accurately. Consequently, the model is not only constrained to reproduce the covariance among pairs of indicators but eventually also the couplings among triplets and quadruplets of indicators.
Second, our analysis reveals that  the model parameters inferred from distinct year data turn out to be statistically indistinguishable. Third, and more importantly, our modelling scheme features a significant forecasting accuracy of the future state of a city based on a previous state of it. The quasi-stationarity of the coupling parameters just mentioned, suggests the possibility to describe the evolution of urban macro-economic indicators as the solution of a stochastic differential equation of the Langevin type. Though in the literature cities are often described as {\em out-of-equilibrium systems}~\cite{batty2008size,batty2017cities}, we observe that treating cities as quasi-equilibrium systems allows predicting the temporal evolution of individual cities. To this end, we assume that the vector of indicators obeys the solution of a Langevin equation whose stationary state is given in terms of the Hamiltonian of our model. Interestingly, our model can forecast the next-year vector of urban features {\it despite the model parameters have been inferred from single-year empirical data, i.e., using no information regarding the temporal evolution}.

We believe that the new framework proposed can find a comprehensive application for a better understanding of urban environments and their evolution. Unlike other inference models suffering from the black-box problem, our ME approach offers a more precise interpretation of the effective mutual influence among the different macro-economic indicators in a given country or region.

The outline of the paper is as follows. In the first section, we introduce the data, the relevant observables derived from it and the ME model build using such observables. In the following two sections, we test the stationarity of the model by comparing the parameters inferred in different years, and we derive a discrete model for temporal predictions using the Langevin Equation. In the last section, we use this model to predict the evolution of individual cities in subsequent years, comparing such predictions with those obtained with a model that explicitly uses the temporal correlations present in the data.

\section*{Results}
\subsection*{Correlations of Rescaled Socio-Economic Indicators}
The data considered in our analysis comes from the INSEE (the French Institut National de la Statistique et des \'Etudes \'Economiques)\footnote{https://www.insee.fr/fr/accueil}, for French {\em communes} from 2006 to 2015. We use this data to build $N$ macro-economic indicators representing the job market (jobs in Primary and Secondary Sectors, in the Tertiary and Quaternary sectors, in Commerce, in Public Administration and services, the Employment rate) and some demographics of each commune (fraction of highly educated people, number of immigrants, average salary per hour). 
In the Appendix, section A, we report the code of each INSEE data variable used to build our indicators, as well as the corresponding INSEE dataset used. The indicators used in our analysis are related to demographic or economic aspects of the French population. Nevertheless, the modelling scheme allows for the inclusion of other kinds of indicators as, crime rates, commercial links between cities, migrations between communes. 

We indicate a generic socio-economic indicator as $X_i^{ (\alpha)}$, where $i$ is the index of the indicator, $i=1,\ldots,N$, and $\alpha$ indicates the commune the indicator refers to. We now define the \emph{rescaled indicators}, ${x}_i^{(\alpha)}$, as
\begin{equation}
{x'}_i^{(\alpha)} = \log_{10} (X_i^{(\alpha)}/(X_i^0 P^{a_i}_\alpha))
\label{eq:scaling}
\end{equation}
where $P_\alpha$ is the population in the commune and $a_i$ is the exponent of the scaling law associated to the $i$-th indicator.  $X^0_i$ is the prefactor of the dependence of $X_i$ on the population size, $X_i = X^0_i P^a_i$. Finally, we divide each indicator by its standard deviation $x_i=x'_i/\sigma(x'_i)$. The scaling procedure and the standardisation might harden the readability of the variables themselves, as compared to the standard way to present socio-economic indicators. Nevertheless, this procedure makes all the variables living in similar spaces and allow for the comparison of communes of different sizes. For readability's sake, if a rescaled indicator of a specific commune is exactly $0$, it means that the actual indicator is precisely the average of all the communes with the same population. Similarly, if the rescaled indicator is $1$, then the value of the actual indicator is one standard deviation larger than the average over the communes with the same population. 

Recent studies have focused on several aspects of the standardized indicators, $x_i$. In~\cite{bettencourt2010urban} it is shown how they exhibit fast decaying spatial correlations. In~\cite{arcaute2015constructing,bettencourt2019interpretation,keuschnigg2019scaling} it has been shown how scaling laws by themselves are not sufficient to predict the evolution of cities. 
In this work we are interested in building a probabilistic generative model in terms of the vector of rescaled indicators, $\x=(x_i)_{i=1}^N$. We will call $\P: \mathbb{R}^N\to\mathbb{R}$ the probability distribution defining the model, and $\<\cdot \>_{\P}$ the expectation value according to it. The generative model is required to reproduce the empirical correlations up to the $m-$th order. To do so, we need to estimate the order of the correlations, $m$, that is relevant and sufficient to describe the data, given the uncertainty associated with the database finiteness. We define the empirical $n$-th order tensor of correlations as
\begin{equation}
C^{(n)}_{i_1, \dots, i_n} = \langle x_{i_1}\dots x_{i_n} \rangle_{\textrm{data}},
\label{eq:n_corr}
\end{equation}

\noindent where $\langle \cdot \rangle_{\textrm{data}}$ indicates the empirical average over the communes belonging to the database (i.e., over the index $\alpha$). Also, we refer to $n$-th order tensor of cumulants, $\bar C^{(n)}$. For each order $n$, we have computed the fraction of elements of the tensors $C^{(n)}$ and $\bar C^{(n)}$ that are significantly different from zero given their statistical error. To this end we adopted the bootstrap error, which accounts for the empirical uncertainty induced by the database finiteness (see the Appendix for details), while the statistical significance refers to a Student t-test. The non-significance of the $n$-th order cumulants indicates, at least, that they cannot be significantly measured due to the database finiteness (this is to be expected, especially for large $n$). Consequently, they should not be considered as a sufficient statistics to be reproduced by $\P$ or, in other words, that $m<n$. Conversely, the presence of significantly nonzero values of the cumulant $\bar C^{(n)}$ imply that one should ask the model to reproduce them (i.e., $m\ge n$). If the $n$-th order correlator is nonzero, this does not imply that $m\ge n$, since they could be due explained by lower-order correlations. For example, even for Gaussian data (for which $m=2$), the $4$-th order correlator $C^{(4)}$ is nonzero in general, while the $4$-th order cumulants ($\bar C^{(4)}_{ijkl} =   C^{(4)}_{ijkl} - C^{(2)}_{ij} C^{(2)}_{kl}-C^{(2)}_{ik}C^{(2)}_{jl}-C^{(2)}_{il}C^{(2)}_{jk}$) vanish. 
We observe that ($n=1$) all the averages, $C^{(1)}_{i}$, of the features are consistent $0$ with a $p$-value larger than $0.05$; ($n=2$) $\sim 91\%$ of $2$-point correlations $C^{(2)}_{ij}$ are non-zero ($p < 0.05$); ($n=3$) $\sim 61\%$ of $3$-point correlations, $C^{(3)}_{ijk}$ are non-zero ($p<0.05$). We, hence, conclude that $m\ge 3$. ($n=4$) While  $\sim 63\%$ of $4$-points correlations are non-zero ($p<0.05$), only $\sim 6\%$ of the cumulant components  $\bar C^{(4)}_{ijkl}$ are significantly nonzero ($p<0.05$). We will consequently consider $m=3$. In other words, we will consider $C^{(2)}$ and $C^{(3)}$ as sufficient statistics that the model $\P$ is constrained to reproduce. 

\subsection*{Inference of the Hamiltonian model for cities}
\begin{figure}[h!]
	\centering
	\includegraphics[width=0.8\linewidth]{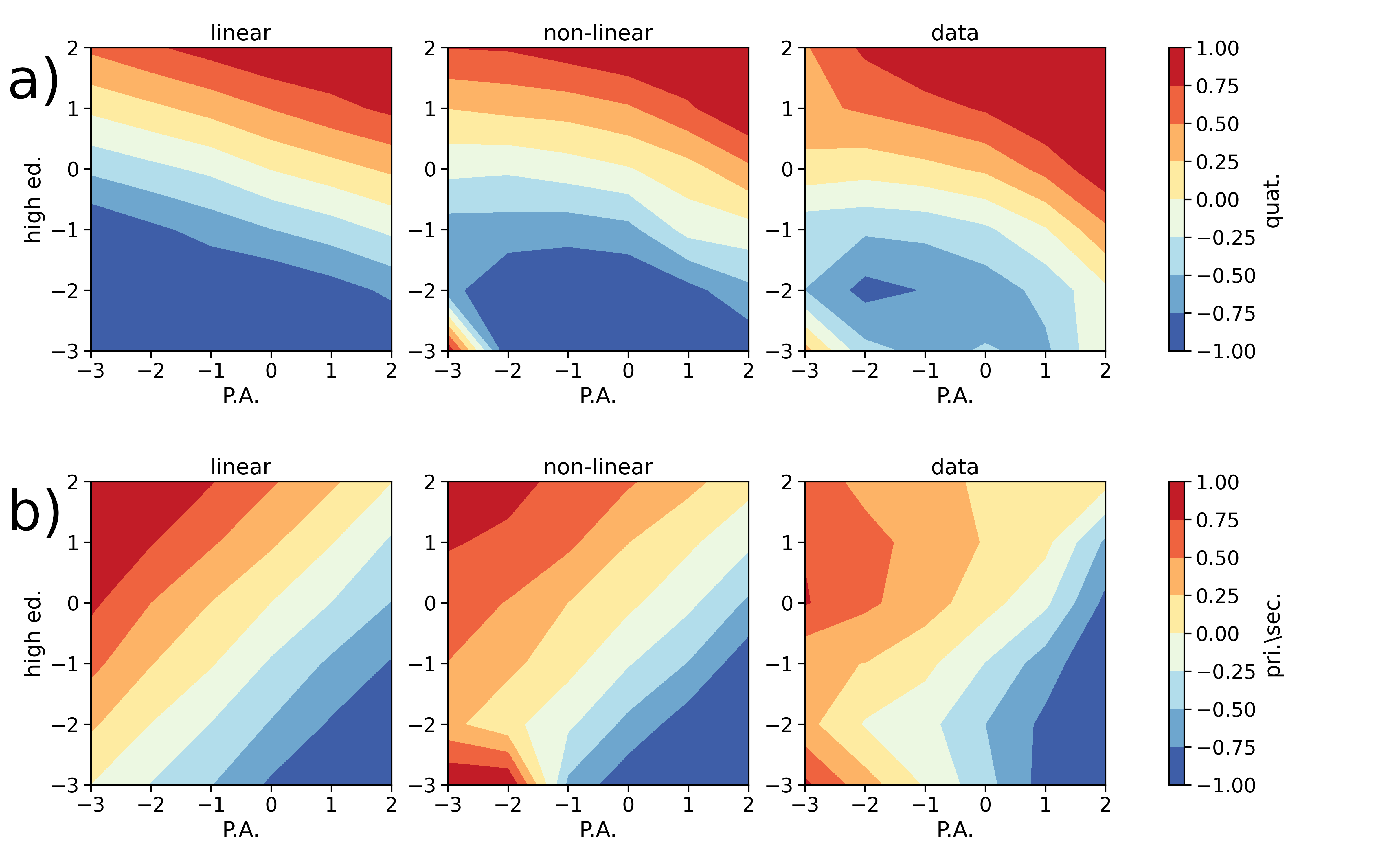}
	\caption{\label{fig:non_linear_prediction} Rescaled indicators for {\rm Jobs in Quaternary Sector} (a) and {\em Jobs in Primary and Secondary Sectors} (b) as functions of the rescaled indicators for {\em  Jobs in Public Administration} (x-axis) and  {\em Fraction of highly educated people} (y-axis).  Areas in red (blue) represent communes with a large (small) value of the rescaled indicator used as dependent variable. The first column (with the label {\em linear}) reports the results obtained with the Hamiltonian model without the terms $J^{(1)}$ and $J^{(3)}$. The second column (with the label {\em non-linear}) reports the results obtained with the complete model of  Eq.~\protect\ref{eq:hamiltonian}. The last column reports the results obtained by binning the points for the communes in the year ($2012$).}
\end{figure}

The ME framework leads to an {\it energy-based} probability distribution (a Maxwell-Boltzmann distribution, in the language of statistical physics), $\P(\x)\propto \exp(-H(\x))$, with an associated Hamiltonian functional $H(\x)$ in the space of socio-economic indicators, whose form is determined by the sufficient statistics:
\begin{equation}
H(\x) = \sum_{ij} J^{(2)}_{ij} x_i x_j + \sum_{ijk} J^{(3)}_{ijk} x_i x_j x_k - \sum_i J^{(1)}_{i} x_i
\label{eq:hamiltonian}
\end{equation}
where $J^(2)$ and $J^(3)$ are the coupling parameters for binary and ternary  interactions, resepctively. The presence of the term with $J^{(1)}$ is required to compensate the effects of the second term with $J^{(3)}$ and make sure  that the averages produced by the model are equal to the empirical ones. The introduction of the $J^{(3)}$ couplings would lead to a non-normalisable distribution $\P(\x)\propto \exp(-H(\x))$ if ${\bf x} \in \mathbb{R}^N$. We have consequently reduced the support of the distribution to the hypercube $x_i \in [-L, L]$ wth $L=6$. In other words, each indicator in the distribution support can be at maximum $6$ standard deviations away from its average. The resulting probability distribution after learning the data exhibits a single absolute maximum in its support, near the origin ${\bf x}={\bf 0}$. Qualitatively, it is a perturbation of the multivariate Gaussian distribution (obtained for null values of the tensor $J^{(3)}$) with respect to which it exhibits larger and asymmetrical tails close to the border of the hypercube (see Appendix section E for a more detailed discussion). The coupling parameters are estimated by the Maximum Likelihood method. Such a maximisation is approximately performed numerically by gradient ascent, simulating a Langevin Dynamics (see Appendix section G) for the estimation of the theoretical correlations appearing in the gradient at each iteration. Once the parameters have been inferred, we perform a convergence and consistency check by verifying the extent to which the model reproduces experimental correlations of $n$-th order. 
\begin{figure}[h!]
	\centering
	\includegraphics[width=0.6\linewidth]{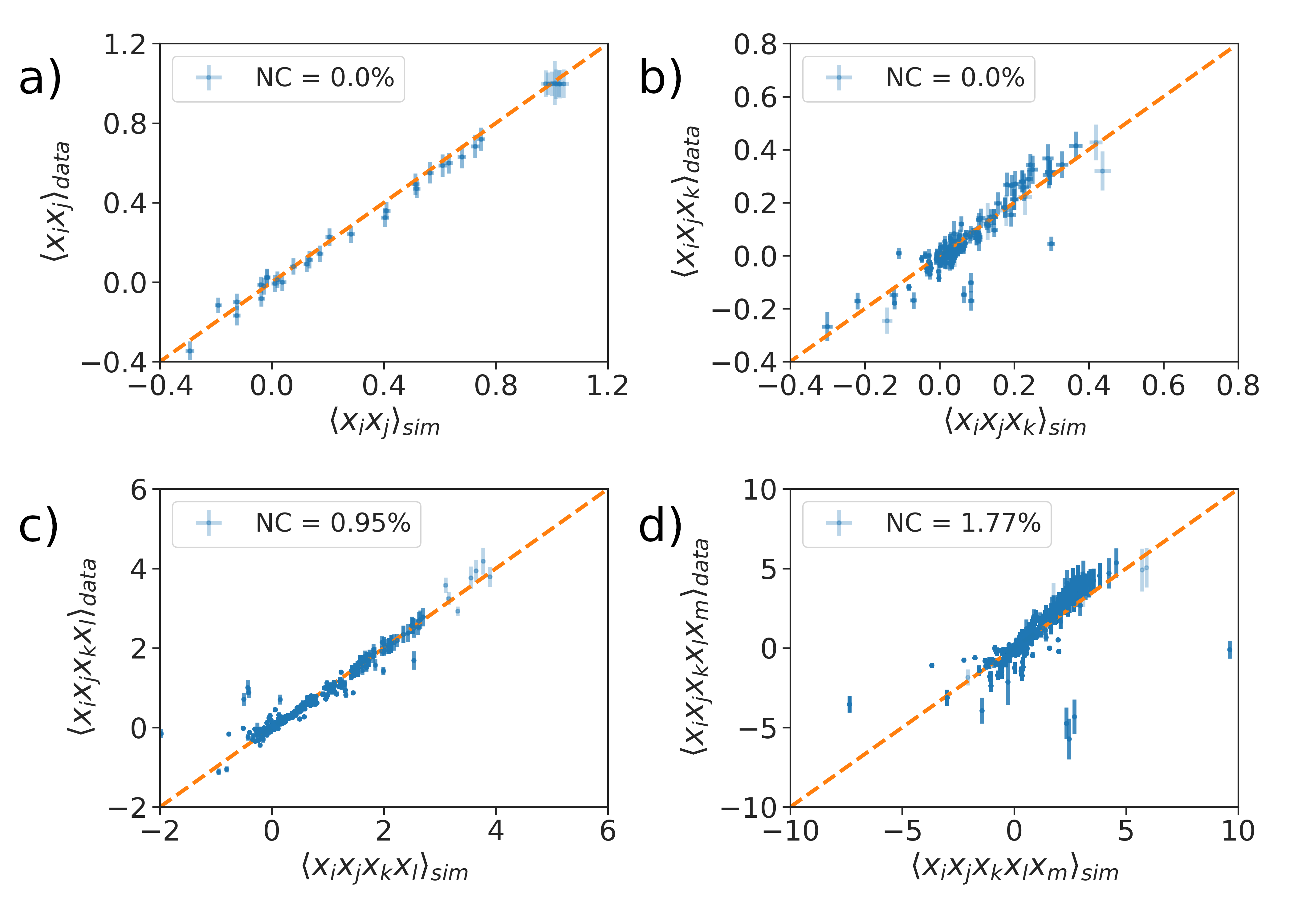}
	\caption{\label{fig:correlations}
		Comparisons between the correlators of order $2$ (a),  $3$ (b), $4$ (c) and $5$ (d)  obtained with the empirical data of year $2012$ ($y$-axis) and the Hamiltonian model~\ref{eq:hamiltonian}($x$-axis). We report the percentage of components of each correlation which is not compatible with the data via a $t$-test with $p$-value $0.05$ (see Appendix section G). 
	}  
\end{figure}
Fig.~\ref{fig:correlations} shows the comparison between the empirical $C^{(n)}$ with those produced by the model taking into account the correlations up to the order $n=5$. The synthetic $C^{(n)}$ have been estimated generating a sample from $\P(\x) \propto \exp (-H(\x))$ (using the Langevin Equation below Eq.~\ref{eq:langevin}), which allows also for an estimation of the standard deviation of each component of $C^{(n)}$ (see G section F). We can use this error, together with the bootstrapped error of the empirical $C^{(n)}$, to perform a t-test of consistency. The percentage of non-compatible components for each $C^{(n)}$ is less than $5\%$, the few discrepancies typically occurring for points with large estimation errors, especially for $n=4$ and $n=5$. This results validates the numerical gradient ascent procedure. Furthermore, the model consistently reproduces correlators at orders $n=4$ and $n= 5$, which were not supposed to be reproduced by construction. The ensemble to these results justifies and validates {\it a posteriori} the Maximum Entropy method and the sufficient statistics used, i.e., using cumulants up to $m=3$.  
\noindent In order to further  validate the above statement, we compared the performances of the model described by Eq.~\ref{eq:hamiltonian} with a simpler Gaussian one in which we removed all terms except the $J^{(2)}$ one:
\begin{equation}
H_G(\x) = \sum_{ij} J^{(2)}_{ij} x_i x_j 
\label{eq:hamiltonian_gaussian}
\end{equation}
The experiment aims to evaluate which model better grasps the interplay between different socio-economic indicators. To illustrate the results, we show, without loss of generality, the dependence of one indicator as a function of the other two. Fig.~\ref{fig:non_linear_prediction} reports the comparison of the models' predictions with the empirical data. In particular, we show the dependence of one rescaled indicator (i.e., {\rm Jobs in Quaternary Sector} in panel (a) and {\em Jobs in Primary and Secondary Sectors}  in panel (b)) on other two rescaled indicators, namely {\em  Jobs in Public Administration} (x-axis) and  {\em Fraction of highly educated people} (y-axis).

\noindent It is evident that the introduction of the non-linear term $J^{(3)}$ increases the model ability to predict, and it is key to capture the non-linear effects present in the data. Given a generative model $\P(\x)$, it is possible to study this dependency by sampling from the conditional probability $\P(x_i | x_j, x_k)$ for several values of $(x_j, x_k)$, being $x_i$ the considered dependent variable. The first column of  Fig.~\ref{fig:non_linear_prediction}, shows this dependency in the case of a model inferred without the terms $J^{(1)}$ and $J^{(3)}$, a simple Gaussian model leading to linear dependencies among all the variables. The prediction of this model is not in agreement with the data, shown in the last column of Fig.~\ref{fig:non_linear_prediction}. Instead, the inclusion of $J^{(1)}$ and $J^{(3)}$ (central panels)  leads to more adherence to the empirical data. Considering, for example, the upper right corner of all the panels, this represents communes with a large number of jobs in public administration and a large number of residents with high education. In this area, both the linear and non-linear models agree with the data, predicting a large number of Jobs in Quaternary (the rescaled indicator is around $1$) and an average number of Jobs in the Primary and Secondary sectors (the rescaled indicator is around $0$). Moving instead to the lower-left corner of each panel, this represents communes with very few Jobs in Public Administration and residents with high-education. In this area, the linear model would predict an average number of Jobs in the Primary and Secondary and very few Jobs in Quaternary (the rescaled indicator is $-3$). However, the data show that in this corner, there is a large number of jobs in both sectors. This fact can be easily explained by the presence in our dataset of mainly industrial areas, poorly served by public administrations and with a scarcely educated population in which workers of every skill commute to work. This behaviour is correctly predicted by the model in Eq.~\protect\ref{eq:hamiltonian}. Other examples, similar to that in Fig.~\ref{fig:non_linear_prediction}, can be found in section E of Appendinx, fully confirming this picture.
\FloatBarrier
\subsection*{Stationarity of the model}
A very important question to ask is whether the modelling scheme proposed in the section above, and synthesized by Eq.~\protect\ref{eq:hamiltonian}, is robust with respect to the empirical data gathered in  different years. In other words, are the values of the interaction parameters $J^{(n)}$ stable when inferred from different years' data? The answer is yes as we illustrate below. 

\begin{figure}[h!]
	\centering
	\includegraphics[width=0.8\linewidth]{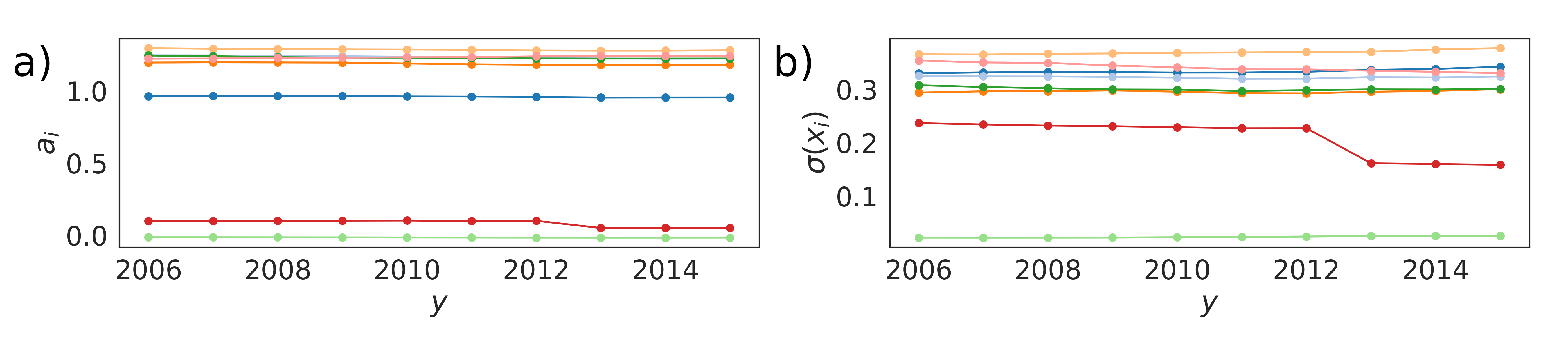}
	\caption{ \label{fig:scaling_sigma_in_time} (a) Scaling exponents $a_i$ for the socio-economic indicators $i$ as a function of time. (b) $\sigma(x_i)$ for the socio-economic indicators $i$  as a function of time. Each indicator is represented by a different color.}
\end{figure}

First, we observe from Fig.~\ref{fig:scaling_sigma_in_time} that most of the exponents of the scaling law, $a_i$, used to define the re-scaled indicators through Eq.~\ref{eq:scaling} are constant in time, and it is so also for the standard deviations $\sigma(x_i)$. Small deviations are only seen for $2$ indicators.

\begin{figure}[h!]
	\centering
	\includegraphics[width=0.8\linewidth]{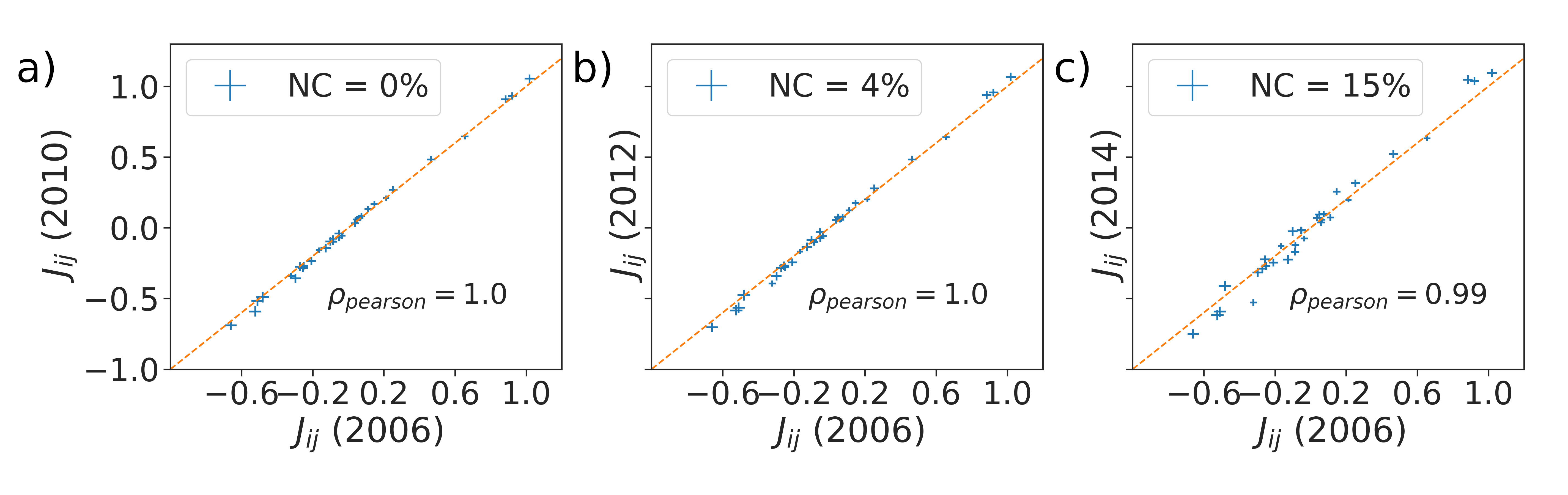}
	\caption{\label{fig:J_comparisons} Comparisons between the $J^{(2)}$ parameters of different years (2006-2010 in panel (a), 2006-2012 in panel (b), 2006-2014 in panel (c)). The dotted line represent the relation of equality (i.e., the diagonal of the first half-plane in the Cartesian space. The percentage of non-compatible (NC) reported refers to the percentage of components that cannot be considered as equal with a $t$-test and a threshold $p$-value of $0.05$.} 
\end{figure}

Having checked the stationarity of the quantities used to define the variables $x_i$, we have addressed the stationarity of the inferred interaction parameters $J^{(n)}$. This is done through a significance t-test (see Appendix section F) of the compatibility between $J^{(1)}(y_1)$, $J^{(2)}(y_1)$, $J^{(3)}(y_1)$ inferred in a certain year $y_1$, and $J^{(1)}(y_2)$, $J^{(2)}(y_2)$, $J^{(3)}(y_2)$ in year $y_2$. Indeed, the parameters are statistically compatible across different years ($p$-value$<0.05$, see Fig.~\ref{fig:J_comparisons} and the Appendix section F for the comparisons of $J^{(1)}$ and $J^{(3)}$). Hence, despite the moderate variation in the scaling exponents and the standard deviations of the indicators, the model is stationary from one year to another. This result does not imply that the indicators of a single commune, $\x^{(\alpha)}$ are not evolving in time. The value of $\x^{(\alpha)}$ are, in fact, {\it non stationary} from one year to another. Instead, the correlations among several indicators, $C^{(n)}$,  stay constant. This remarkable result paves the way for a description of the evolution of socio-economic indicators in terms either of {\it equilibrium models}, in a statistical-physical sense, or out-of-equilibrium stationary models~\cite{chou2011non}.

\subsection*{Forecasting the time evolution of the socio-economic indicators}

\begin{figure}
	\centering
	\includegraphics[width=0.95\linewidth]{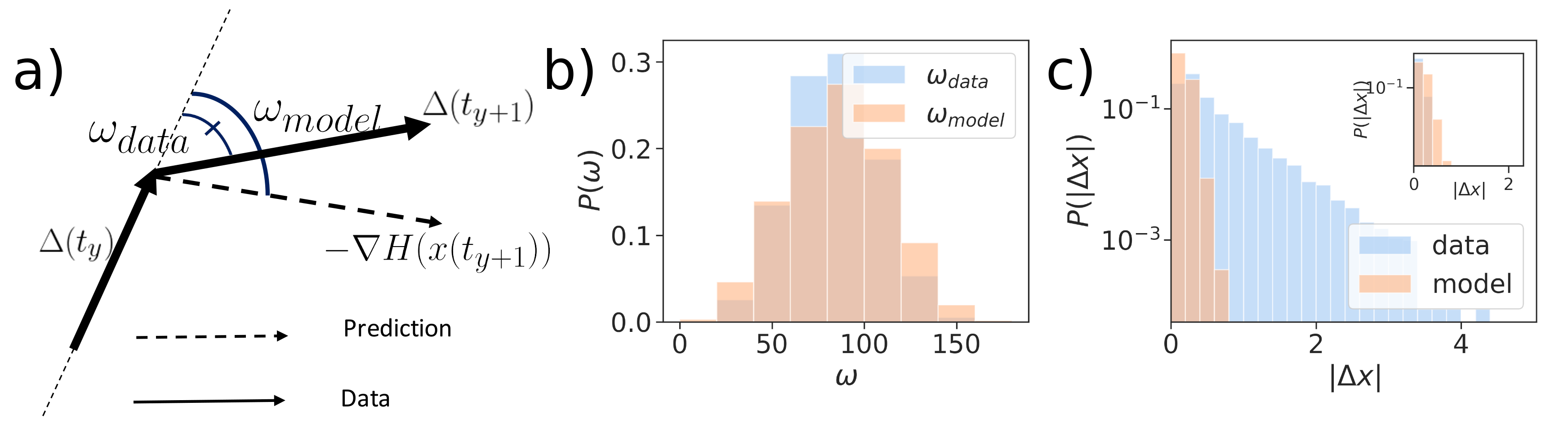}
	\caption{(a) Graphical representation of the angles $\omega_{\textit{data}}$ (the smaller angle in panel a) and $\omega_{\textit{model}}$ (the larger angle in panel a) , identified respectively by the variation $\bm\Delta(t_{y+1})$ at time $t_y$ and the predicted variation of the model at the same time $-\nabla H (\x(t_y))$, and by the two subsequent variations $\bm\Delta(t_y)$ and $\bm\Delta(t_{y+1})$. (b) Comparison between the angle $\omega_{\textit{data}}$ between the velocity of the system at consecutive times and the angle $\omega_{\textit{model}}$ between the velocity of the system and the velocity predicted by Eq.~(\ref{eq:langevin}). (c)  Variations $\Delta x_i(t_y)$ for all the components $i$ of the feature vector in the same two cases. In the inset, we show the same comparison excluding communes with a population larger than $10^4$.}
	\label{fig:equilibrium_hypothesis}
\end{figure}

\begin{figure}
	\centering
	\includegraphics[width=0.95\linewidth]{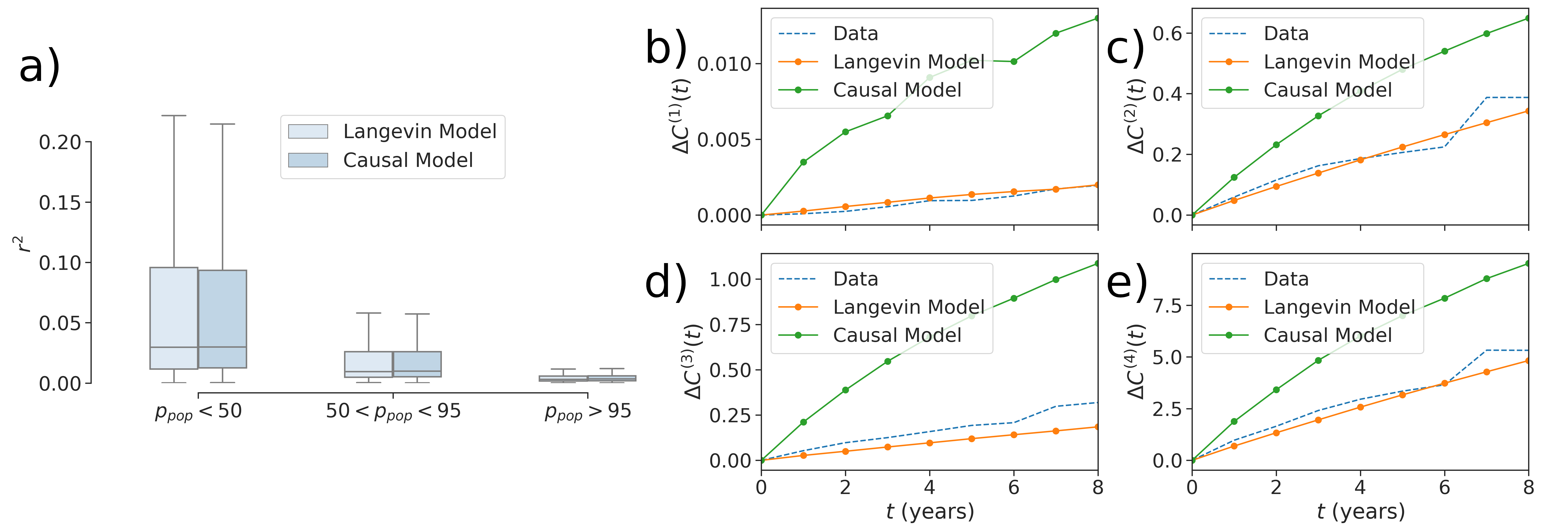}
	\caption{(a) Distribution of $r^2$ computed according to equation (\ref{eq:r}) and with a Causal Inference model, inferred using temporal information explicitly. The values of $r^2$ have been divided according to the percentile of the commune population distribution. (c-b-d-e) Evolution of the distance between macroscopic observables computed in different years with respect to those computed in $2012$ ($t=0$ on the x-axis). }
	\label{fig:predictionbysize}
\end{figure}

In this section we test the forecasting capacity of our Hamiltonian modelling scheme, i.e. the ability to predict future values of the rescaled indicators starting from a given starting condition. To this end, we consider the Langevin equation for the stochastic temporal evolution of a vector field~\cite{gardiner2004handbook}. This provides a simple model for the continuous-time dynamics of vector $\x$, whose stationary distribution is our generative model, $\P(\x) \propto \exp(-H(\x))$:
\begin{equation}
\frac{d\x(t')}{dt'}(t)= -\nabla H (\x(t)) + \bm\eta(t),
\label{eq:langevin}
\end{equation}
where $\bm\eta(t)$ is $N$-dimensional vector of independent random variables with vanishing average, extracted from a probability distribution $h$, satisfying $\langle {\eta_i}(t) \eta_j(t') \rangle_h  = \delta(t - t')\delta_{ij}$. This is a strong assumption, that implies not only the stationarity of the distribution of $\x$ in the large $t$-limit, but also {\it thermal equilibrium} (or, roughly speaking, absence of probability currents) \cite{chou2011non}. Nevertheless, this assumption might still be useful to make predictions about the trajectories of individual cities.
The distribution of $\bm{\eta}$'s, $h(\bm\eta)$ can be chosen arbitrary. We use a Laplacian noise i.e., $h( \bm{\eta}) \propto \exp( - \vert \bm\eta \vert /2)$ (see Appendix section G). In equation (\ref{eq:langevin}) the time is a continuous variable, whose physical interpretation is not straightforward for our model, since our data is defined in discrete time. We will indicate different years with a specific intrinsic time $t_y$ so that the consecutive year time is $t_{y+1}$. Approximating the derivative by a finite difference, $dt = t_{y+1} - t_{y}$, we can derive the probability of observing the feature vector $\x(t_{y+1})$ for a certain city after having observed the values of the previous year $\x(t_y)$ as:
\begin{equation}
p_{dt}(\x(t_y+dt) | \x(t_y) )= \prod_{j=1}^N \sqrt{\frac{1}{2dt}} \exp \left (  -\frac{ \sqrt{2} \left| x_j(t_y+dt) - x_j(t_y) - f_j(\x(t_y))\, dt\right|}{\sqrt{dt}} \right),
\label{eq:small_gauss_jump}
\end{equation}
where $f_j(\x(t_y)) = -\frac{\partial H}{\partial x_j} (\x(t_y))$. Assuming that our system is governed by Eq.~(\ref{eq:langevin}), we can use Maximum Likelihood to estimate the value of $dt$ that best reproduces the transitions between subsequent years (see Appendix section G).
According to the discrete-time Langevin model, the variation of a feature vector from one year to the next, $\bm\Delta(t_y) = \x(t_{y+1}) - \x(t_y) $ should be proportional to minus the gradient of the Hamiltonian $H$ plus some Laplacian noise. Hence, such variation should be on average parallel and proportional to $-\nabla H(\x_y)$ (where $\x_y=\x(t_y)$).
To check this hypothesis, we compare the angle $\omega_{\textit{data}}$ between two consecutive variations of the feature vector  $\bm\Delta(t_y)$  and $\bm\Delta(t_{y+1})$, with the angle $\omega_{\textit{model}}$ between $-\nabla H (\x(t_y))$ and $\bm\Delta(t_y)$. Fig.~\ref{fig:equilibrium_hypothesis} (a) shows this comparison. We can interpret $\omega_{\textit{data}}$ as the angle between two consecutive velocities of the system, while $\omega_{\textit{model}}$ is the angle between the velocity at time $t_y$ and the predicted velocity at time $t_{y+1}$ (i.e.,  $-\nabla H(\x(t_{y+1}))dt$). The agreement between the angles indicates that the rotation of the velocity at different times is compatible with that predicted by the Langevin dynamics.
The remarkable agreement between the data and the synthetic sample justifies equation (\ref{eq:langevin}) as a model of the evolution of the urban indicators. A comparison between the modules of each $\bm\Delta(t_y)$ obtained with data and with simulations is shown in Fig.~\ref{fig:equilibrium_hypothesis} (c). In this case the agreement is less strong, since the real data distribution is broader for extreme values. However, if we restrict the comparison only to large communes, with a population $P>10^4$, the agreement increases (inset of Fig.~\ref{fig:equilibrium_hypothesis}, c), suggesting that the discrete-time Langevin approach, Eq.~(\ref{eq:langevin}), is, at least, a good model for the evolution of large cities. 
This fact emerges also from a further analysis of the model dynamical forecasting accuracy. We have performed a statistical test to evaluate the accuracy of the model prediction for $\x(t_{y+1})$ from the real data $\x(t_y)$, according to Eq. (\ref{eq:small_gauss_jump}). 
For every city and every year $y$, we consider the quantity:
\begin{equation}
r^2_y = \frac{1}{N} \sum_{j=1}^N | x_j(t_{y+1}) -  x_j(t_y) - f_j(t_y)\, dt |^2,
\label{eq:r}
\end{equation}
\noindent i.e., the average square residual between the actual feature vector at time $t_{y+1}$, $\x(t_{y+1})$, and the average model prediction. i.e., $\x(t_y) - {\bf f}(t_y)dt$.
Fig.~\ref{fig:predictionbysize} (a) shows the distribution of the $r_y^2$ variables after dividing the sample according to the percentile of population of each city. The model performs better as the size of the city increases, in agreement with the results in Fig.~\ref{fig:equilibrium_hypothesis}.  In order to have an accuracy baseline,  Fig.~\ref{fig:equilibrium_hypothesis} shows the same  $r_y^2$  divided according to the population for a Causal Inference (CI) model \cite{presse2013principles} in which the information about temporal correlations of the indicators at consecutive and non-consecutive years are explicitly inferred by Maximum Likelihood. Both distributions are statistically compatible. It is strikingly surprising that the accuracy of both models are equivalent, since the discrete-time Langevin model {\it has been inferred from the single-year data}, hence {\it neglecting the  database information regarding the time evolution}. Yet, our model forecasts rather accurately. In the CI the information regarding the temporal evolution is {\it inferred} from the data. In our discrete-time Langevin model it is, instead, {\it postulated} through Eq. (\ref{eq:small_gauss_jump}), and does not need to be inferred. This however comes at the cost of introducing a non-stationary term in the model, absent in the discrete-time Langevin dynamics. We can show this computing  the observables $C^{(n)}(y)$ with $n=1,2,3,4$ from equation (2) in the main text for each year $y$ of our data. For each $n$, we define $\Delta C^{(n)}(y) = \vert C^{(n)}(y) - C^{(n)}(0) \vert_2$  where $\vert \cdot \vert_2$ is the Frobenius Norm. This quantity indicates for each year, how much the observable $C^{(n)}(y)$ has shifted from its initial value $C^{(n)}(0)$.  Thus, we produce two synthetic samples obtained by making each commune in the first year of our data evolve according to the Langevin Equation and to the CI model. By computing the observables $C^{(n)}(y)$ for each time step of the two synthetic samples, we can compute the corresponding value of $d^{(n)}(y)_{\textit{Langevin}}$ and $d^{(n)}(y)_{\textit{Causal}}$. Fig.~\ref{fig:predictionbysize} panels  (b), (c), (d), (e) shows $d^{(n)}(y)$ as a function of $y$ for the data, the Langeving Model and the Causal Model. We see that the data shows a small shift in the observables and the Langeving model is always more coherent the Causal Model in reproducing it.
\section*{Discussion}
In relatively recent times, the phenomenon of urbanisation is proceeding at an unprecedented pace. Nowadays, urban environments represent the pumping heart of modern-day life with all its diverse aspects affecting progress and innovation. Despite the importance of the phenomenon, little is known about the critical determinants of cities and their evolution. It has been observed that socio-economic indicators related to urban environments follow scaling laws with the city size. Those regularities helped to formulate hypotheses about the deep meaning of the observed self-similarity as well as the mechanisms for the emergence of these laws. One of the big successes of scaling theory applied to city science is the possibility it opens to define re-scaled socio-economic indicators, which, on their turn, allow for comparing different cities at different population scales. Despite these successes, scaling theory represents a {\em a posteriori} description of cities, and little can be said {\em a priori} about the dynamical evolution of these relevant entities and their constituents. However, a modelling framework is still lacking in the science of cities that, through a careful description of the interactions and the couplings among the diverse aspects of the urban fabric, could allow us to assess the status of cities and create validated scenarios of future evolution.
With this paper, we made a step forward in this direction by proposing a first Maximum Entropy generative model of towns based on careful observation of modern cities as witnessed by data of French ``communes'' in the period from 2006 to 2015. This generative model defines an Hamiltonian written in terms of a vector of socio-economic indicators whose coupling parameters are inferred through an unsupervised Maximum Likelihood approach. The Hamiltonian defines a probabilistic model that takes into account non-linear effective interactions up to the order $m=3$ (i.e., it takes into account couplings of two and three socio-economic indicators). In this way, our approach goes beyond a principal component analysis, and it allows reproducing the non-linear correlations observed in the empirical data up to higher orders four and five. We show that the inclusion of these couplings is necessary in order to correctly describe the data, which exhibit in many cases behaviours that are far from linear. The whole approach allows for projecting cities in a high-dimensional landscape (defined in terms of the socio-economic features) where each existing town sits in a specific spot, and its dynamics occurs along the manifold defined by the Hamiltonian model. 
Interestingly, the inferred model is quite robust, and the different coupling parameters turn to be invariant over several years. Along with the stationarity of the scaling laws, this result suggests that the statistical laws governing the socio-economic indicators can be approximately considered as constants in time. If we adopt the terminology of stochastic processes, the stationarity of the inferred model allows for a description of the dynamics of an individual city in terms of stationary out-of-equilibrium or a quasi-equilibrium model. Following the latter and most straightforward approach, we have proposed a dynamical model based on the Langevin equation, compatible with our inferred Boltzmann probability distribution (defined by an effective Hamiltonian function), and assessed its predictive power. More in detail, we made specific predictions about the status of a city at time $t+1$, knowing its status at time $t$. Surprisingly, the forecasting accuracy of such a dynamical model is generally quite good, and it grows with the population size of the considered commune.
Our results pave the way for a novel and precise, yet interpretable, predictive modelling of urban environments from a macro-economic point of view. Our framework is also suited to be applied to a causal inference of the effects of shocks, stress conditions or exogenous events, and to model the recovery of cities after them. This whole framework could help to forecast the decline or growth of towns and shed light on the causes of such behaviours. For example, a variation in the model parameters could model the effects of changing the national and international scenario on the urban system, as well as the impact of policies in the job market. %
\appendix
\section{APPENDIX}

\section{INSEE Data about ``Communes'' in France}
\begin{figure}
	\centering
	\includegraphics[width=0.5\columnwidth]{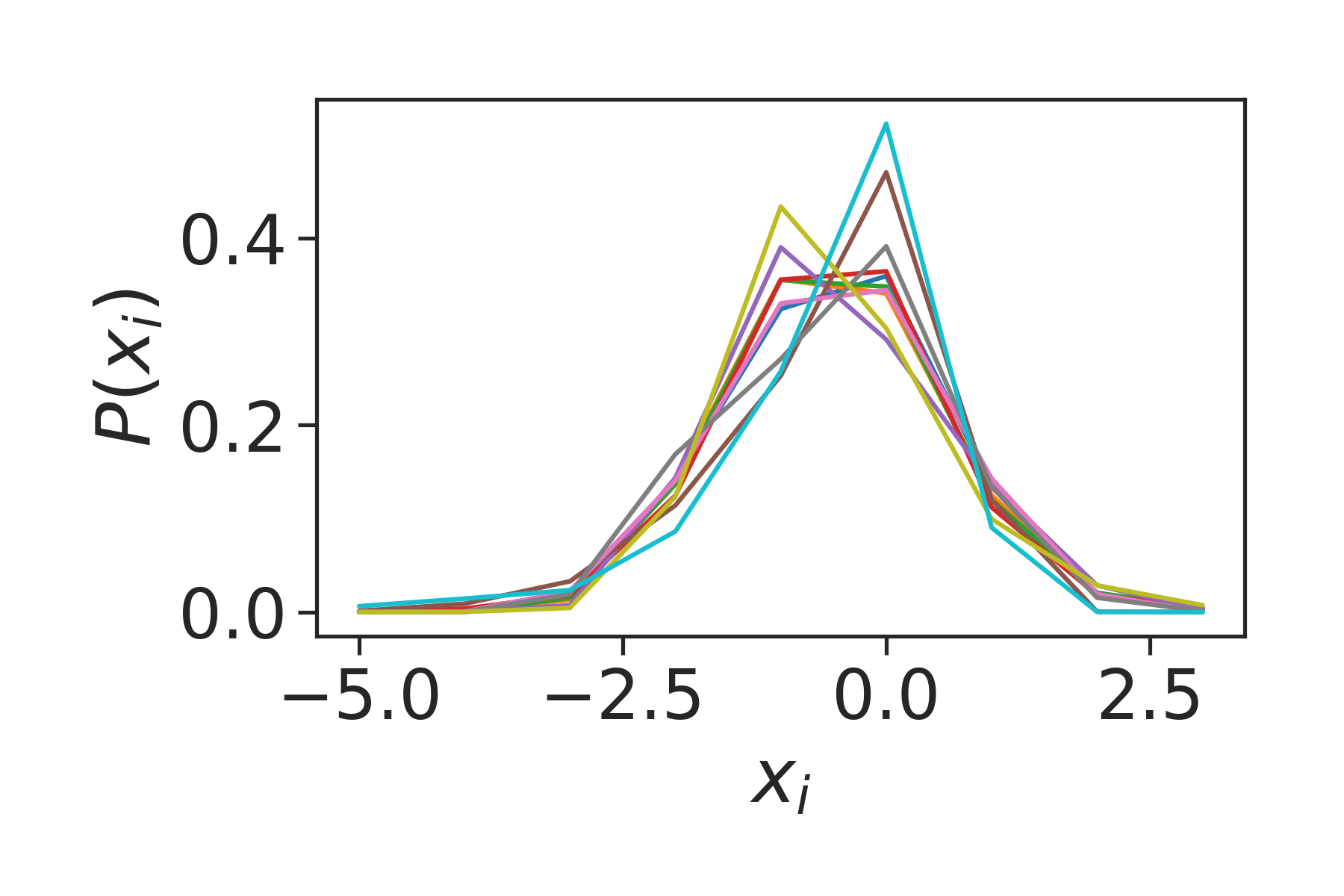}
	\caption{Distributions for the re-scaled $x_i$ variables for the year $2012$. The other years are not shown, but similar results can be found in those cases.}
	\label{figsi:distributions}
\end{figure}

The data we consider in this work comes from the French Institut National de la Statistique et des \'Etudes \'Economiques (INSEE) and can be downloaded freely from its website (\url{https://www.insee.fr/fr/accueil}). The data we downloaded concerns several aspects of the French ``Communes'' which are the smallest administrative units in the country, ranging from few hundreds of inhabitants to several millions. In our analysis we arbitrarily removed all the administrative units with less than $20$ inhabitants. There is much information in the data we downloaded that has not been used in the work, while other has been aggregated to obtain $10$ socio-economic indicators representing some aspects of the job market and of the population. INSEE provides yearly snapshots of data about the communes. In our work, we downloaded data from 2006 to 2015, from different data sources. In the following we indicate with $\{Y\}$ a variable which is the last two digits of each year (e.g. $\{Y\}=12$ in $2012$).
\newline
From "Emploi - Population active'' data we built the variables
\begin{enumerate}
	\setcounter{enumi}{0}
	\item {\bf Jobs in Primary and Secondary Sectors}: the sum of the variables  C$\{Y\}$\_EMPLT\_CS1 (agriculture operators), C$\{Y\}$\_EMPLT\_CS6 (factory workers), C$\{Y\}$\_EMPLT\_AGRI (workers in agriculture), C$\{Y\}$\_EMPLT\_INDUS (employed in industry),  C$\{Y\}$\_EMPLT\_CONST (workers in construction).
	\item {\bf Jobs in the Tertiary Sector}: C$\{Y\}$\_EMPLT\_CS4 (intermediary professions)
	\item {\bf Jobs in Commerce}: the sum of C$\{Y\}$\_EMPLT\_CTS (workers in commerce) and C$\{Y\}$\_EMPLT\_CS2 (works in artisan's shops)
	\item {\bf Jobs in Quaternary}: C$\{Y\}$\_EMPLT\_CS3 (workers in intellectually superior jobs)
	\item {\bf Jobs in Public Administration and services}: C$\{Y\}$\_EMPLT\_APESAS (workers in public administration, teaching, health institutes and social aid).
	\item {\bf Employment rate}: ratio between P$\{Y\}$\_ACTOCC1564 (active employed population)  and P$\{Y\}$\_ACT1564 (active population).z
\end{enumerate}
From "Dipl\^omes - Formation'' data we built the variables:
\begin{enumerate}
	\setcounter{enumi}{6}
	\item {\bf Fraction of highly educated}: ratio between P$\{Y\}$\_NSCOL15P\_SUP (Population not in school more than $15$ years old with higher education degrees) and P$\{Y\}$\_NSCOL15P (Population not in school more than $15$ years old)
\end{enumerate}
From "Population par sexe, \^age et situation quant \'a l'immigration''
\begin{enumerate}
	\setcounter{enumi}{7}
	\item {\bf Number of immigrants}:  sum of AGE400\_IMMI1\_SEXE1,
	AGE400\_IMMI1\_SEXE2, AGE415\_IMMI1\_SEXE1, AGE415\_IMMI1\_SEXE2, AGE425\_IMMI1\_SEXE1, AGE425\_IMMI1\_SEXE2, AGE455\_IMMI1\_SEXE1, AGE455\_IMMI1\_SEXE2. Here SEXE1 indicates males and SEXE2 indicates females. Moreover, AGE400 indicates population less than $15$ years old, AGE415 population of age between $15$ and $24$, AGE425  population of age between $24$ and $54$, AGE455 population more than $54$ years old.
\end{enumerate}
From "Salaires et revenus d'activit\'e" data:
\begin{enumerate}
	\setcounter{enumi}{8}
	\item {\bf Average salary per hour}: the variable SNHM$\{Y\}$.
\end{enumerate}
Finally the population of each commune can be read from the "Evolution et structure de la population'' data, in the P$\{Y\}$\_POP variable.
Each variable $X_i$ has been found to be dependent on the population $P$ via the power-law relation $X_i = X^0_i P^a_i$. The exponents $a_i$ for each variable in each year are shown in the main text and are found to be roughly constant in time. We use this relation to define the variable $x_i = \log_{10} (X_i/(X_i^0 P^{a_i}))$, which then we re-scale by their standard deviation $\sigma(x_i)$. In this way, we obtain variable which are bell-shaped with the same variance as shown in Fig.~\ref{figsi:distributions}.
\FloatBarrier
\section{Error Estimation with Bootstrapping and $t$-test}
To perform the $t$-tests in the main text we need to estimate the error on our observables $C^{(n)}$. Considering a certain function $f(x)$ defined on our data $\{ x^\alpha \}_{\alpha=1}^{N_c}$, we can easily estimate its average over the sample using
\begin{equation}
\langle f \rangle_{\textit{data}} = \frac{1}{N_c} \sum_\alpha f(x^\alpha).
\label{eqsi:avgdata}
\end{equation}
In order to assign an error to the average we can divide our sample in $M$ sub-samples of $0.8N_c$ elements, built by randomly picking elements of $\{ x^\alpha \}_{\alpha=1}^{N_c}$ (with repetitions). We can then use (\ref{eqsi:avgdata}) on each sub-sample, finding a set of mean values $\{ \langle f \rangle_{m} \}_{m=1}^{M}$ (bootstrap sample) where $m$ identifies different sub-samples. The average over this set can the be assumed as an estimate of $\langle f \rangle_{\textit{data}}$. Similarly, the standard deviation over the set $\{ \langle f \rangle_{m} \}_{m=1}^{M}$  can be assumed as standard error. We indicate this two quantities with $\bar{f}$ and $\sigma(f)$ respectively.
We can use the set $\{ \langle f \rangle_{m} \}_{m=1}^{M}$ to perform a $t$-test to check the compatibility of $\langle f \rangle_{\textit{data}}$ with a certain value $\mu_f$, via the statistics
\begin{equation}
t_{\textit{1sample}} = \frac{\bar{f} - \mu_f}{\sigma(f)}.
\end{equation}
This statistics is used to perform a double-tailed test over the $t$-distribution with $M-1$ degrees of freedom under the null hypothesis that $\bar{f}$ is different from $\mu_f$. We reject this hypothesis if the $p$-value of the test is larger than $0.05$.
In case we need to compare two empirical averages (e.g. when we compared the components of $C^{(4)}$ and those of $(C_{\textit{gauss}})^{(4)}$), we build a bootstrap sample for each quantity. Identifying these quantities with $f$ and $g$ respectively and with $M_f$ and $M_g$ the dimension of the bootstrapped sample, we use the statistics
\begin{equation}
t_{\textit{2sample}} = \frac{\bar{f} - \bar{f}}{\sqrt{\sigma(f)^2 + \sigma(g)^2}}.
\end{equation}
to perform a double tailed test with a $t$ distribution with 
\begin{equation}
\nu = \frac{\sigma(f)^2 + \sigma(g)^2}{\frac{\sigma(f)^4}{M_f-1} + \frac{\sigma(g)^4}{M_g-1}},
\end{equation}
degrees of freedom under the null hypothesis that $\bar{f}$ and $\bar{g}$ are different. We can use this test also to compare empirical averages with those produced by the model exchanging the bootstrapped average and standard error with those obtained with a Langevin simulation (in which case the size of the sample is the number of simulation steps).
\FloatBarrier
\section{Maximum Entropy and Parameters Estimation}
Let's consider a data set of $N_c$ points that can be considered several realizations of the same distribution $\{ x^\alpha \}_{\alpha=1}^{N_c}$. Each $x^\alpha \in \mathcal{R}^N$ and we indicate with $x^\alpha_i$ its $i$-th component. Suppose we have identified a set of observables $O_\lambda(x)$ with $\lambda$ an integer index, which are function of $\mathcal{R}^N$ and are relevant for the description of our dataset. Maximum Entropy (ME)\cite{martyushev2006maximum} provides an interesting framework for deriving a generative model which preserves the average of the observables measured with the data, $\langle O_\lambda \rangle_{\textit{data}}$. In ME the goal is to find a distribution $\P(x)$ maximising its entropy under the constraints that the average of the observables computed with $\P(x)$ should be the same as in the data. In other words, in ME we have to maximize the functional:
\begin{equation}
\Gamma[ \P ] =  S[ \P ] + \sum_\lambda J_\lambda\; ( \langle O_\lambda \rangle_{\textit{data}} - \langle O_\lambda \rangle_{\P}),
\label{eqsi:maxent}
\end{equation}
where $S[ \P ] = -\int dx \P(x) \log \P(x)$ is the Entropy of the distribution $\P$ and $\langle  f \rangle_{\P} = \int dx f(x)\P(x)$ is the average of the function $f$ over the distribution $\P$. In other words, equation (\ref{eqsi:maxent}) is the Lagrangian function which maximises the entropy under the constraints that the observables produced by $\P$ should be the same as those in the data. Hence, $J_\lambda$ are the Lagrange multipliers related to each constraint. With some straightforward calculations, we can show that maximizing equation (\ref{eqsi:maxent}) with respect to $\P$, is equivalent to maximize the loglikelihood
\begin{equation}
\mathcal{L}(J_\lambda) = \frac{1}{N_c} \sum_\alpha \log \P(x^\alpha;J_\lambda),
\label{eqsi:logli}
\end{equation}
with respect to $J_\lambda$, where $\P$ is defined as
\begin{equation}
\P(x) = \frac{1}{Z} \exp \left ( - \sum_\lambda J_\lambda O_\lambda(x) \right).
\label{eqsi:prob}
\end{equation}
In equation (\ref{eqsi:prob}) $Z$ the ``partition function'' well-known in Statistical Physics and $H(x) = \sum_\lambda J_\lambda O_\lambda(x)$ is the Hamiltonian function of the system.
It is possible to show that maximizing equation (\ref{eqsi:logli}) equals to solve the equations:
\begin{equation}
\frac{\partial \mathcal{L}}{\partial J_\lambda} = \langle O_\lambda \rangle_{\P} - \langle O_\lambda \rangle_{\textit{data}} = 0.
\label{eqsi:gradients}
\end{equation}
However, this would require to know a closed form for $\langle O_\lambda \rangle_{\P}$ which is typically not the case. Another common approach to estimate the maximum of the likelihood function is that to perform a gradient ascent using equations (\ref{eqsi:gradients}). The problem with $\langle O_\lambda \rangle_{\P}$ at each step of the ascent is solved by using Langevin simulations to compute these averages and then use that to compute the gradients. This is the approach we have used for our  system. Note that typically this is not feasible if the phase space becomes too big. However, in our case $N=9$ allows to have estimates of $\langle O_\lambda \rangle_{\P}$ with reasonably short simulations.
%
%
%
%
%
\FloatBarrier
\section{Maximum Entropy for rescaled socio-economic indicators}
\begin{figure}
 \centering
 \includegraphics[width=0.5\columnwidth]{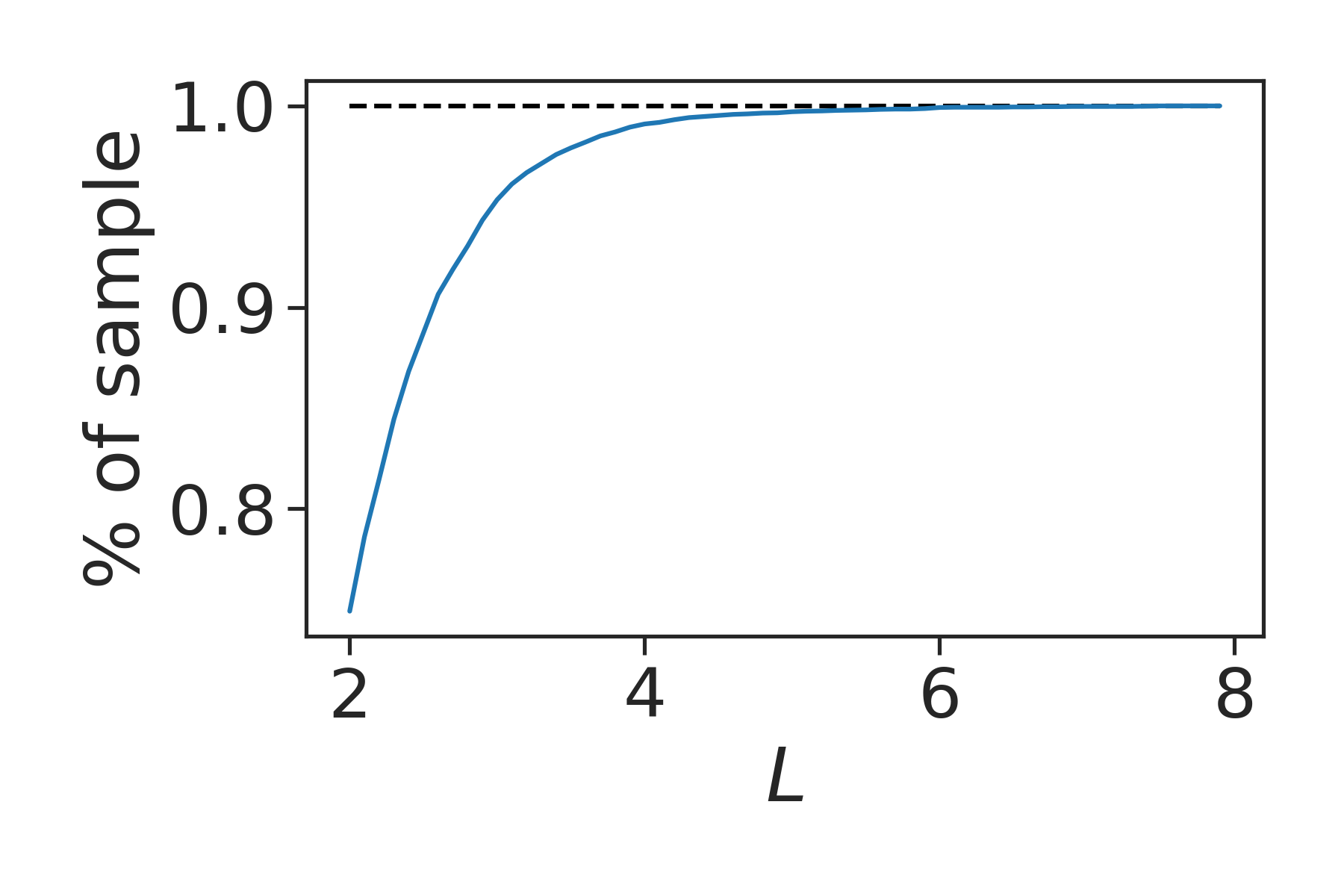}
 \label{figsi:sample_cube}
 \caption{  Percentage of data points contained withing the hypercube $I = [-L, L]^N$ as a function of $L$.} 
\end{figure}

Considering the data in the main text, we are interested in estimating the effective interactions between the re-scaled indicators $x_i$. In the main text we have identified some observables related to the correlations between the indicators. In particular, we have seen that besides $C^{(2)}_{i,j} = \langle x_{i}x_{j} \rangle_{\textit{data}}$ also $C^{(3)}_{i,j,k} = \langle x_{i} x_{j} x_k \rangle_{\textit{data}}$ cannot be considered equal to $0$. If we assume $C^{(2)}_{i,j}$ as the only relevant observables, according to the framework defined in the previous paragraph we would end up with a model 
\begin{equation}
    \P_0(x) \propto \exp \left ( -\sum_{ij} J^{(2)}_{ij}x_i x_j \right ),
    \label{eqsi:gaussprob}
\end{equation}
i.e. a Gaussian model which is not capable of producing correlations $C^{(n)}$ with odd $n$. The fact that $C^{(3)}_{i,j,k}$ cannot be considered $0$ forces us to assume it as a relevant observable to be put in the model. The inclusion of $C^{(3)}$ might lead to $C^{(1)}$ different from $0$ which is instead observed in the data. Thus, we will include $C^{(1)}_{i}=0$ for every $i$ as an observable in the model. We obtain the model defined in the main text in which there is a contribution to the Hamiltonian of $3$-points interactions
\begin{equation}
    \P(x) \propto \exp \left ( -\sum_{ij} J^{(2)}_{ij}x_i x_j  - \sum_{ijk} J^{(3)}_{ijk}x_i x_j x_k  + \sum_{i} J^{(1)}_{i} x_i \right).
    \label{eqsi:3prob}
\end{equation}
The introduction of the term $J^{(3)}$ in the Hamiltonian is such that the distribution $\P(x)$ cannot be normalised if its support is $\mathbb{R}^N$. In other words, there will be directions in $\mathbb{R}^N$ that will make the distribution grow indefinitely. However, there might be values of the coupling parameters that will allow for some local maxima that will constrain the dynamics of the system for a finite time, before it diverges for $t\rightarrow \infty$. To prevent this behaviour, we can bound the system around these maxima redefining the support of the probability $\P(\x)$ as $I = [-L, L]^N$, i.e., a hypercube centred on the origin. The choice of the value $L$ influences the model training and efficiency in a non-trivial way. If $L$ is too small, some parts of the space that are populated by the empirical data could be excluded. For sufficiently large values of $L$, the function may develop, during training, a global maximum in $I$ different from the convex, perturbed Gaussian maximum near the origin. Fig.~\ref{figsi:sample_cube} shows the percentage of data points within the hypercube as a function of $L$. We see that the first value that almost all the sample if $L>5$, hence we choose $L=6$, i.e. $6$ standard deviations of the sample.

To estimate the Lagrange multipliers $J^{(2)}_{ij}$ and $J^{(3)}_{ijk}$, we need to find the values maximizing (\ref{eqsi:prob}) via gradient ascent. This requires to be able to compute exactly the log-likelihood $Z$ to be computed. Estimating $Z$ is quite a hard task typically. To circumvent this problem, we will use an approach widely used for training Energy Based models in Machine Learning. We can write the log-likelihood of our model as:
\begin{equation}
    \mathcal{L} = - \langle H(\x) \rangle_{\textit{data}} - \log Z,
\end{equation}
Where $\langle \cdot \rangle_{\textit{data}}$ indicates the average on the sample data. Taking the gradient of the above expression we find 
\begin{equation}
    \nabla \mathcal{L} = -\langle \nabla H(\x) \rangle_{\textit{data}} + \langle \nabla H(x) \rangle_{\P},
    \label{eqsi:approxgrad}
\end{equation}
where $\langle \cdot \rangle_{\P}$ is the average for the model. This average can be approximated at each training step by averaging over a sample obtained with numerical simulations (e.g. by iterating equation (\ref{eqsi:langevin_discrete}). If we use this approximation we can see from equation (\ref{eqsi:approxgrad}) that maximising the log-likelihood is equivalent to optimise the cost function:
\begin{equation}
    \mathcal{C} = -\langle H(x) \rangle_{\textit{data}} + \langle H(x) \rangle_{\P},
    \label{eqsi:approxcost}
\end{equation}
that we can use to monitor the development of the training. %

%
%
%
%

\begin{figure}
    \centering
    \includegraphics[width=0.75\columnwidth]{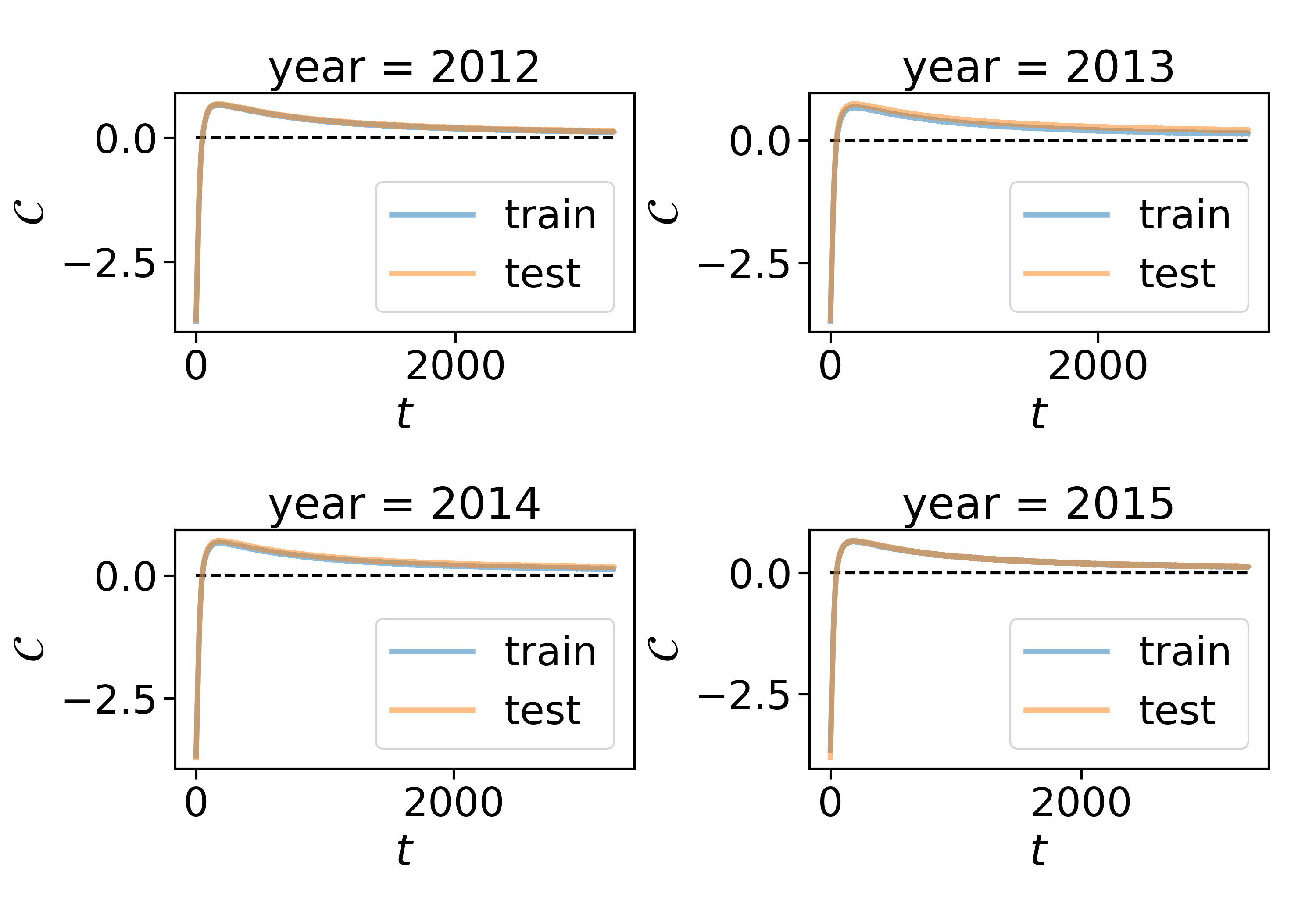}
    \caption{ Cost function for the train and test data during training as a function of the training step.} 
    \label{figsi:logli_test}
\end{figure}

\begin{figure}[!h] 
 \centering
 \includegraphics[width=0.9\columnwidth]{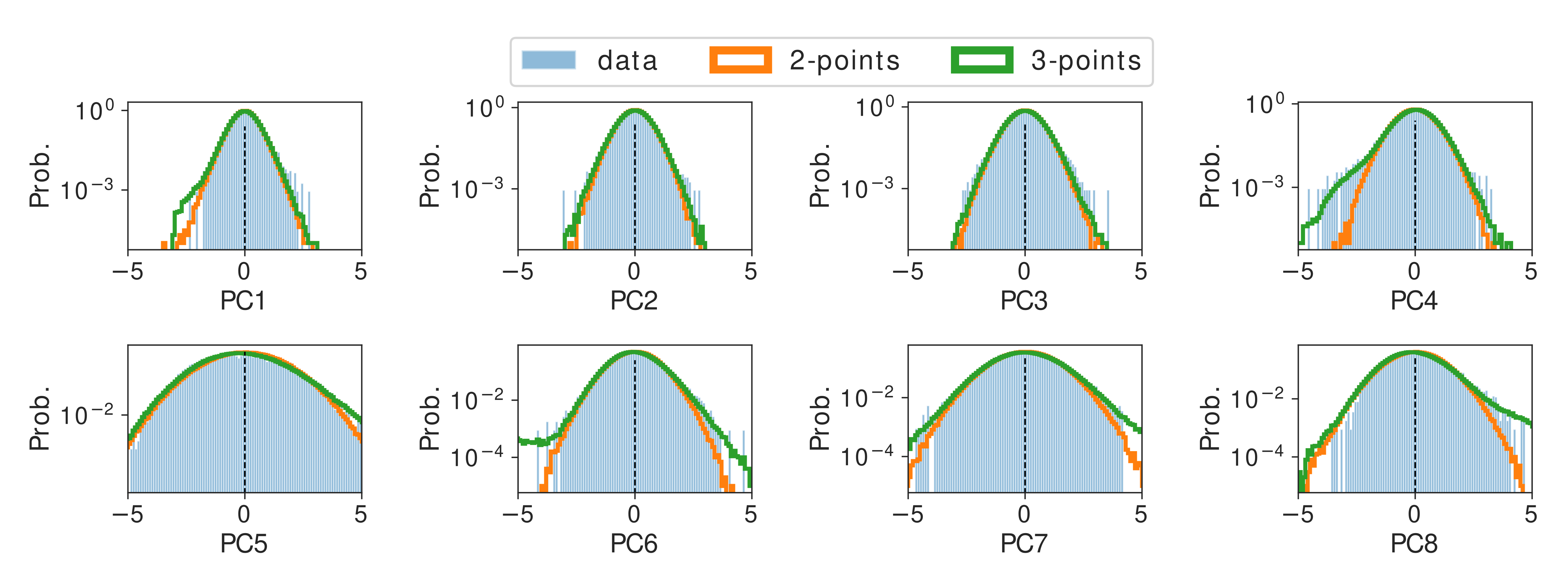}
 \label{figsi:PC_distr}
 \caption{ =Distribution of the Pricipal Components of $J^{(2)}$ obtained with the data (blue bars), the Gaussian model of equation (\ref{eqsi:gaussprob}) (orange line) and with the model with 3-points interactions and bounded support (\ref{eqsi:3prob}) (green line).}
\end{figure}

To avoid over-fitting when estimating the parameters of the model, we divided the sample in a training set ($\approx 70\%$ of the whole sample) and test set (the remaining part). In order to make the two samples as similar as possible, we initially divided the whole sample in percentiles of the population distribution: from the $0^{th}$ percentile to the $5^{th}$; from the $5^{th}$ percentile to the $10^{th}$ and so one. We divided each classes in training and test with the proportion of 
$70\%$ and $30\%$, having in this way a global train and test sample with the same population distribution. This was done in order to not over-represent small cities which are more numerous that the large ones. \newline
The algorithm used to estimate $J^{(2)}_{ij}$ and $J^{(3)}_{ijk}$ is then:
\begin{enumerate}
    \item Starting at $t=0$, we set $J^{(2)}_{ij} = 1/2 \delta_{ij}$, $J^{(3)}_{ijk} = 0$ and $J^{(1)}_{i} = 0$, which is equivalent to a system of non-interactive variables with variance equal to $1$. We estimate the starting value of the cost function (\ref{eqsi:approxcost}) for the training and test data.
    \item At each time step, we generate a sample from the current version of $\P(x)$, iterating equation (\ref{eqsi:langevin_discrete}) with $dt=0.1$ for at least $10^6$ steps. To prevent the simulations from diverging, we bound the dynamics to the box $I = [-6,6]^N$.
    \item We use the generated sample and the training data to estimate the gradients
        \begin{equation}
            \begin{split}
                \frac{\partial \mathcal{L}}{\partial J^{(1)}_{i} } &= \langle x_i \rangle_{\textit{data}} - \langle x_i \rangle_{\P},\\
                \frac{\partial \mathcal{L}}{\partial J^{(2)}_{ij} } &= \langle x_i x_j \rangle_{\P} - \langle x_i x_j \rangle_{\textit{data}},\\
                \frac{\partial \mathcal{L}}{\partial J^{(3)}_{ijk} } &= \langle x_i x_j x_k\rangle_{\P} - \langle x_i x_j x_k \rangle_{\textit{data}}\\
            \end{split}
        \end{equation}
    \item We update the parameters using
        \begin{equation}
            \begin{split}
                J^{(1)}_{i} &\leftarrow J^{(1)}_{i} + \eta_J^{(1)} \frac{\partial \mathcal{L}}{\partial J^{(1)}_{i} } ,\\
                J^{(2)}_{ij} &\leftarrow J^{(2)}_{ij} + \eta_J^{(2)} \frac{\partial \mathcal{L}}{\partial J^{(2)}_{ij} } ,\\
                J^{(3)}_{ijk} &\leftarrow J^{(3)}_{ijk} + \eta_J^{(3)}\frac{\partial \mathcal{L}}{\partial J^{(3)}_{ijk} }\\
            \end{split}
        \end{equation}     
    \item We compute the new value of the cost function (\ref{eqsi:approxcost}) for the training and test data and we update $t$ by $1$.
    \item We restart from point $2$ until the test log-likelihood stops growing.
\end{enumerate}
The perturbative form for $Z$ used to estimate the log-likelihood requires the contribution of $J^{(3)}$ to be smaller than that of $J^{(2)}$. Hence, we set $\eta_J^{(2)} = 0.01$ and $\eta_J^{(1)}=\eta_J^{(3)}=0.001$.  In Fig.~\ref{figsi:logli_test} we show the cost function curves for all the train and test data for four different years. After a maximum, the cost function decreases, approaching monotonically zero for large values of the number of iterations. The curves for the training and test sets are indistinguishable for some years, or present non-significant differences. We conclude that there are not overfitting issues: the model generalises well to non-observed data. \newline We can study how the probability distribution (\ref{eqsi:3prob}) defined in the hypercube differs from a Multivariate Gaussian distribution (\ref{eqsi:gaussprob}). We will assess the impact of the tensor $J^{(3)}$ in the distribution through a comparison with the Gaussian model in equation (\ref{eqsi:gaussprob}), in terms of the principal components (PCs), or the projections of the physical variables ${\bf x}$ on the eigenvectors of the $J^{(2)}$ matrix. If we do so, the model becomes a set of non-interacting Gaussian models. Fig.~\ref{figsi:PC_distr} shows the comparison among the linear model (a collection of independent normal distributions over the PCs), the non-linear model and the empirical data, as a function of the first PCs. We can see that on each PC the model (\ref{eqsi:gaussprob}) predicts a Gaussian distribution centred in $0$ (orange line), qualitatively similar to the empirical distribution (blue bars). Doing the same for the model in equation (\ref{eqsi:3prob}) gives a slightly different result. We can see in Fig.~\ref{figsi:PC_distr} that the introduction of a bounded dominion allows the model to reproduce the empirical bell-shaped distribution (green line). However, the introduction of $J^{(3)}$ modifies the shape distribution tails, which are now more adherent to the empirical one.

\FloatBarrier
\section{Other Examples of Prediction of a Dependent Variable}
\begin{figure}
	\centering
	\includegraphics[width=0.95\columnwidth]{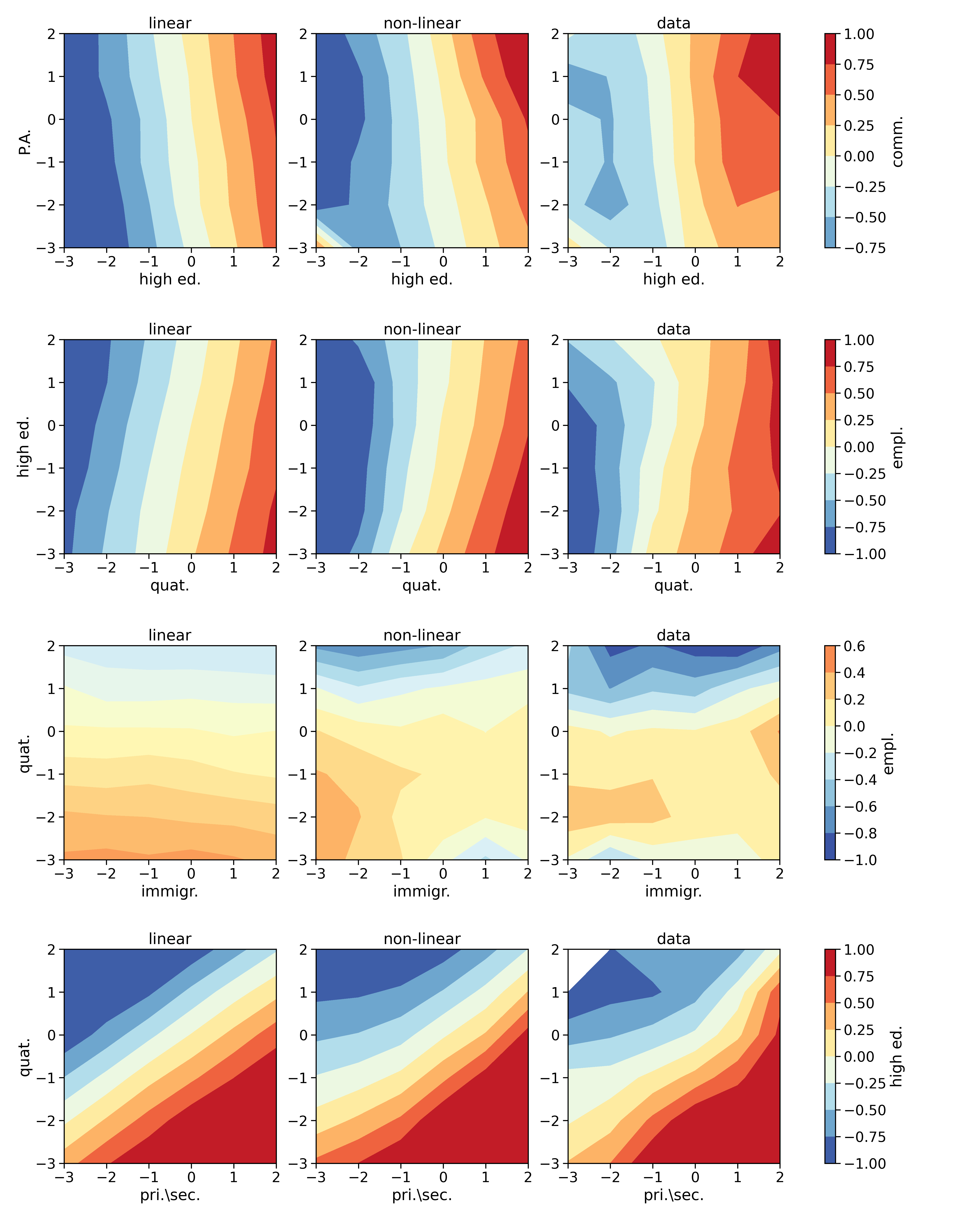}
	\caption{ Some rescaled indicators (indicated in the label of the colorbar) as functions of the some couples of rescaled indicators (indicated on the x-axis and the y-axis of each panel).  Areas in red (blue)represents communes with a large(small) value of the rescaled indicator used as dependent variable. The first column are the results obtained with the Hamiltonian model without the terms $J^{(1)}$ and $J^{(3)}$. The second column are the results obtained with the complete model including those terms. The last column of panels are the results obtained by binning the points for the communes in the year ($2012$).}
	\label{figsi:comparisons}
\end{figure}

\begin{figure}[h!]
	\centering
	\includegraphics[width=0.95\linewidth]{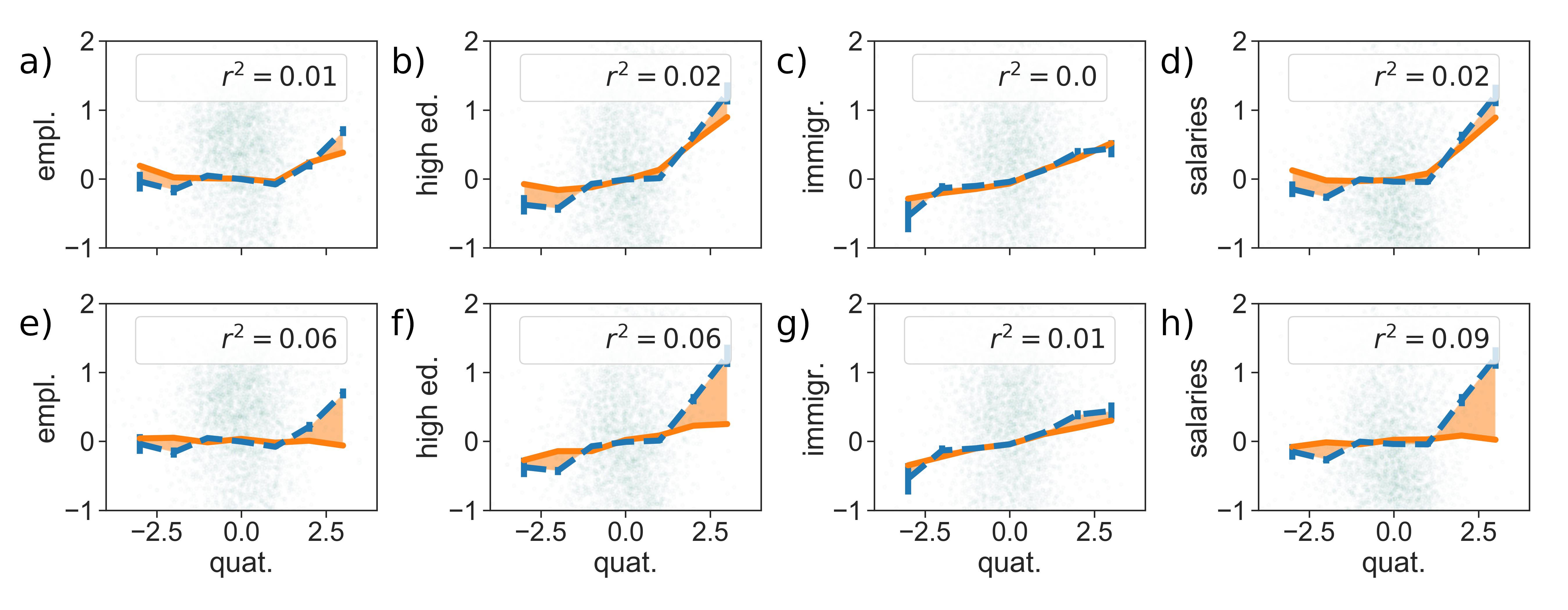}
	\caption{\label{sifig:non_linear_prediction_old} Rescaled indicators for {\em Employment rate}, {\em Fraction of highly educated people}, {\em Number of immigrants}, {\em Average salary per hour } indicators as a function of the rescaled indicator for number of jobs in the quaternary sector as measured with the empirical data (dashed blue line) and with two different Hamiltonian model, the complete one (upper row of panels) of model in the maix text and the simple Gaussian one (lower row of panels) without the terms $J^{(1)}$ and $J^{(3)}$ (orange lines) . Green points represent all the points in the data-set for the considered year ($2012$). The mean squared errors between the data and the models are also shown.}
\end{figure}

In the main text we have shown some examples of predictions of the model, when an indicator is chosen as a dependent variable and another two are used as the independent ones. In Fig.~\ref{figsi:comparisons} , we show some other examples for the $\P_0$ (only $C^{(2)}$ correlations are use in the model) and the $\P$ (also $C^{(1)}$ and $C^{(3)}$).  We can see that the pattern observed in the main text is reproduced for almost every one of the shown cases, i.e. there is more agreement between of the model and the data if $J^{(3)}$ interactions are taken into account.  Considering the fist and second columns of panels, the theoretical predictions are obtained as the average of the distributions $\P_0(x_i | x_j, x_k)$ and $\P(x_i | x_j, x_k)$, where $x_i$ is the chosen dependent variable and $x_j$ and $x_k$ are the two chosen independent ones.  Sampling from these distributions can be made by means simulating the corresponding Langevin dynamics     (\ref{eqsi:langevin_discrete}), keeping fixed the variables $x_j$ and $x_k$  at each step.
In a similar, but more simple fashion we can study the dependence of just one indicator with respect to another. In Fig.~\ref{sifig:non_linear_prediction_old} , we show  four indicators - the employment rate, the fraction of highly educated people, the number the average salary per hour - as a function of the the number of jobs in the quaternary sector. 
It is evident that the introduction of the nonlinear term $J^{(3)}$ increases the model predictive ability, and it is key to capture the non-linear effects present in the data also in this more simple case.  Let us take for instance panels (b) and (f) of Fig.~\ref{sifig:non_linear_prediction_old} reporting the behaviour of the rescaled indicator for the fraction of highly educated citizens as a function of the rescaled indicator for number of jobs in the Quaternary sector . Panel (b) is reporting the comparison of the prediction of the full model with non-linear terms, with the empirical data. Panel (f) shows the same comparison for a simpler Gaussian model described  without $J^{(3)}$ . We can see that in this case, when the rescaled indicators for jobs in the Quaternary sector is smaller than $1$, the increase in the Fraction of Highly Educated citizens is relatively small. In this region, this indicator is always close to $0$, indicating a commune with an average number of highly educated individuals. For values above $1$ (i.e., the number of this jobs in the Quaternary sector is more than $1$ standard deviation to the average of the communes with the same population), the rescaled fraction of highly educated citizens starts to increase more rapidly. The model with only binary interactions fails to predict this behaviour, which is instead well reproduced by the model in with higher-order interactions. This result stays valid for other rescaled indicators as the employment rate (panels (a) and (e) of Fig.~\ref{sifig:non_linear_prediction_old}) and the average yearly salary (panels (d) and (h) of Fig.~\ref{sifig:non_linear_prediction_old}). In the case of the number of immigrants (panels (g) and (c)), no non-linear behaviour is present in the data, and both the models predict the dependence correctly on the rescaled values of the number of jobs in the Quaternary sector. 
\FloatBarrier
\section{Stationarity of the Inferred Models}
\begin{figure}
	\centering
	\includegraphics[width=0.99\columnwidth]{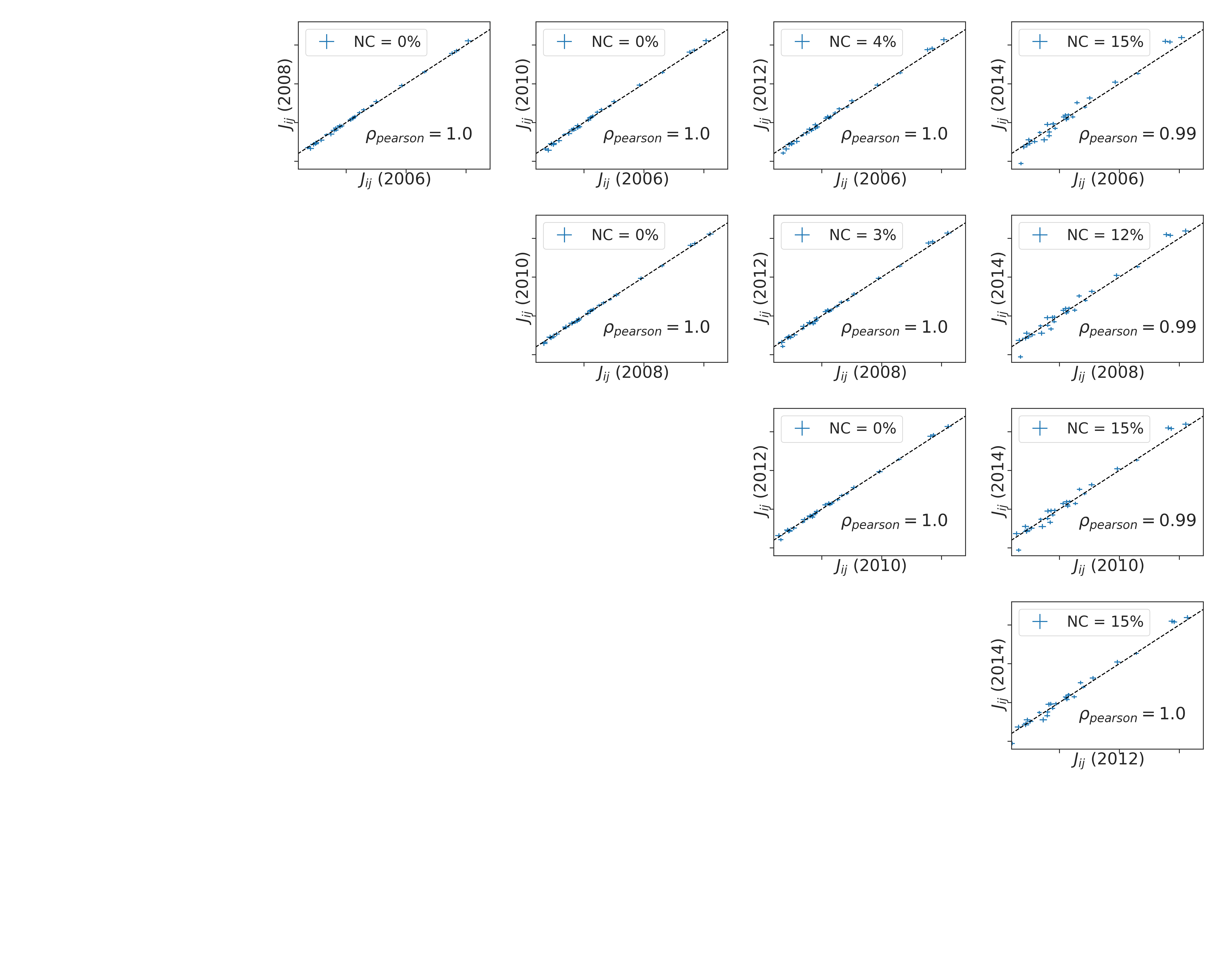}
	\caption{
		\label{figsi:J2_comparisons} Comparisons between the components of $J^{(2)}$ in different years. Errors of each parameter are reported in the plot. Dotted line represents the identity relation. The percentage of component with a $p$-value from $t$-test below $0.05$ is shown in the legend.}
\end{figure}

\begin{figure}
	\centering
	\includegraphics[width=0.99\columnwidth]{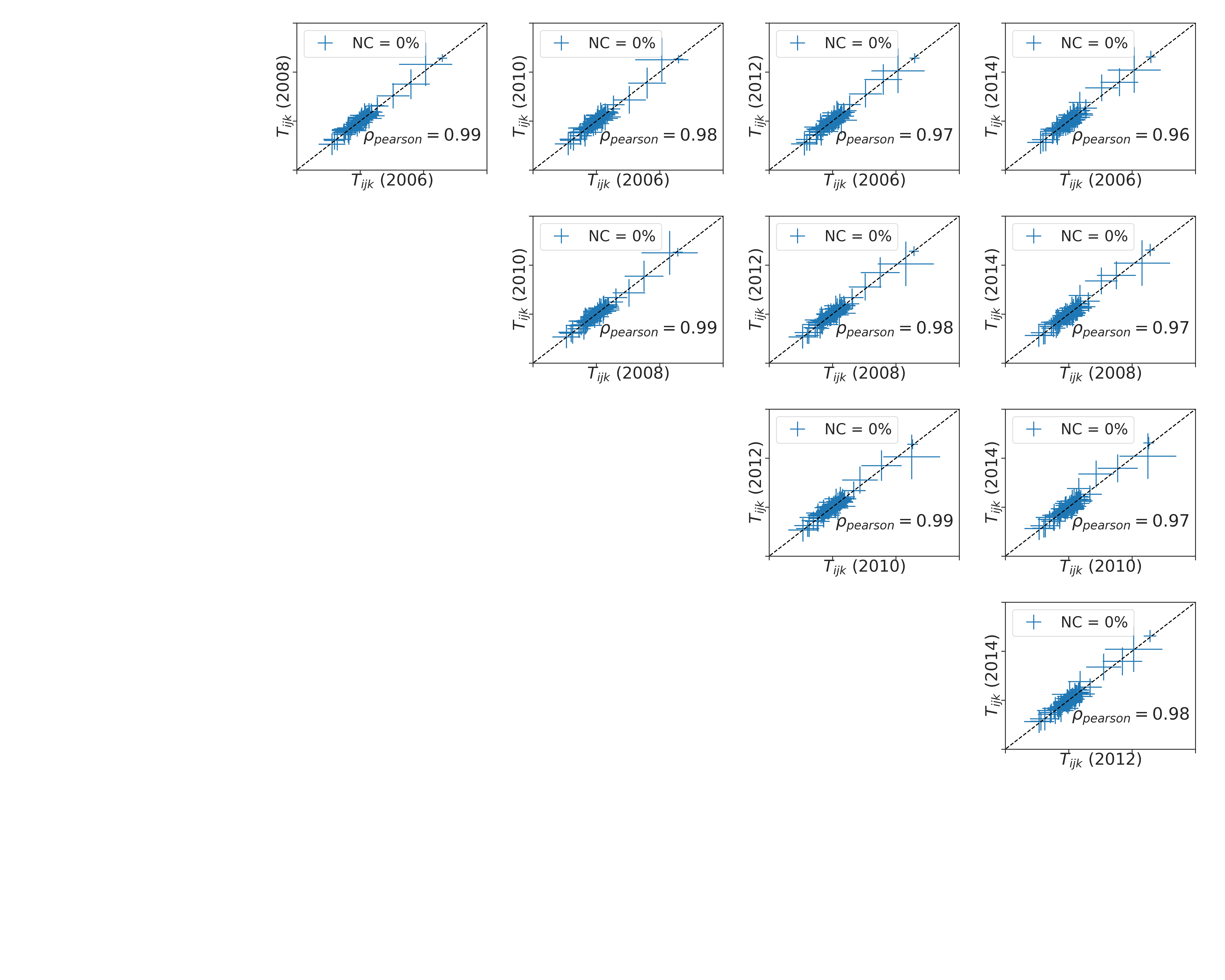}
	\caption{\label{figsi:J3_comparisons}Comparisons between the components of $J^{(3)}$ in different years. Errors of each parameter are reported in the plot. Dotted line represents the identity relation. The percentage of component with a $p$-value from $t$-test below $0.05$ is shown in the legend.}
\end{figure}

Indicating with $J^{(2)}(y_1)$ and $J^{(3)}(y_1)$ the parameters inferred for the data in a certain year $y_1$, it is possible to compare them with those of another year $y_2$. To make statistical comparisons, it is needed to have an idea of the errors associated with each inferred parameter. Errors for the parameters can be computed using the Fisher Information matrix $\mathcal{I}$. In fact, the parameters estimated with Maximum Likelihood can be considered as a coming from a multivariate normal distribution, whose averages are the real parameters and the co-variance matrix is given by the inverse of $\mathcal{I}$. For our system $\mathcal{I}$ is defined as:
\begin{equation}
\begin{split}
\mathcal{I}(J^{(2)}_{ij},J^{(2)}_{lm}) &= - \frac{\partial}{\partial J^{(2)}_{ij}} \frac{\partial}{\partial J^{(2)}_{lm}} \log Z = C^{(4)}_{ijlm}- C^{(2)}_{ij}C^{(2)}_{lm}\\
\mathcal{I}(J^{(2)}_{ij},J^{(1)}_{k}) &= - \frac{\partial}{\partial J^{(2)}_{ij}} \frac{\partial}{\partial J^{(1)}_{k}} \log Z = C^{(3)}_{ijk}- C^{(2)}_{ij}C^{(1)}_{k}\\
\mathcal{I}(J^{(2)}_{ij},J^{(3)}_{lmn}) &= - \frac{\partial}{\partial J^{(2)}_{ij}} \frac{\partial}{\partial J^{(3)}_{lmn}} \log Z = C^{(5)}_{ijlmn}- C^{(2)}_{ij}C^{(3)}_{lmn}\\        
\mathcal{I}(J^{(3)}_{ijk},J^{(3)}_{lmn}) &= - \frac{\partial}{\partial J^{(3)}_{ijk}} \frac{\partial}{\partial J^{(3)}_{lmn}} \log Z = C^{(6)}_{ijklmn}- C^{(3)}_{ijk}C^{(3)}_{lmn}.\\
\mathcal{I}(J^{(3)}_{ijk},J^{(1)}_{l}) &= - \frac{\partial}{\partial J^{(3)}_{ijk}} \frac{\partial}{\partial J^{(1)}_{l}} \log Z = C^{(4)}_{ijkl}- C^{(3)}_{ijk}C^{(1)}_{l}\\
\mathcal{I}(J^{(1)}_{i},J^{(1)}_{l}) &= - \frac{\partial}{\partial J^{(1)}_{i}} \frac{\partial}{\partial J^{(1)}_{l}} \log Z = C^{(2)}_{il}- C^{(1)}_{i}C^{(1)}_{l}
\end{split}
\end{equation}
To compute $\mathcal{I}$ we generate a sample from $P(x) \propto \exp( -H(x) )$ iterating equation (\ref{eqsi:langevin_discrete}) with $dt=0.1$ for at least $10^6$ steps. We then use the produced sample to estimate the observables $C^{(n)}$. The errors associated to each parameter will be then computed using the corresponding element on the diagonal of $\mathcal{I}^{-1}$ as variance, and in turn using such variance to compute the standard error. As an example, the standard error of the estimate of $J^{(2)}_{ij}$ is given by $\sqrt(\mathcal{I}^{-1}(J^{(2)}_{ij},J^{(2)}_{ij}) / N_c) $, where $N_c$ is the number of points in the training set. 
Once we have computed all the errors for each components of $J^{(1)}(y)$, $J^{(2)}(y)$ and $J^{(3)}(y)$ for each year, we can make $t$-tests for each one of their components with null hypothesis that they are compatible. We reject the null hypothesis if the $p$-value of the test is larger than $0.05$. Fig.~\ref{figsi:J2_comparisons} and Fig.~\ref{figsi:J3_comparisons} show the scatter-plot of the corresponding components of $J^{(1)}$, $J^{(2)}$ and $J^{(3)}$ for different years. Each component is plotted with its error and the percentage of components that have failed the $t$-test are shown in the legend of each plot. We can see from these figures that the parameters are quite similar between different years and typically the hypothesis of compatibility cannot be rejected.
\begin{figure}
	\centering
	\includegraphics[width=0.99\columnwidth]{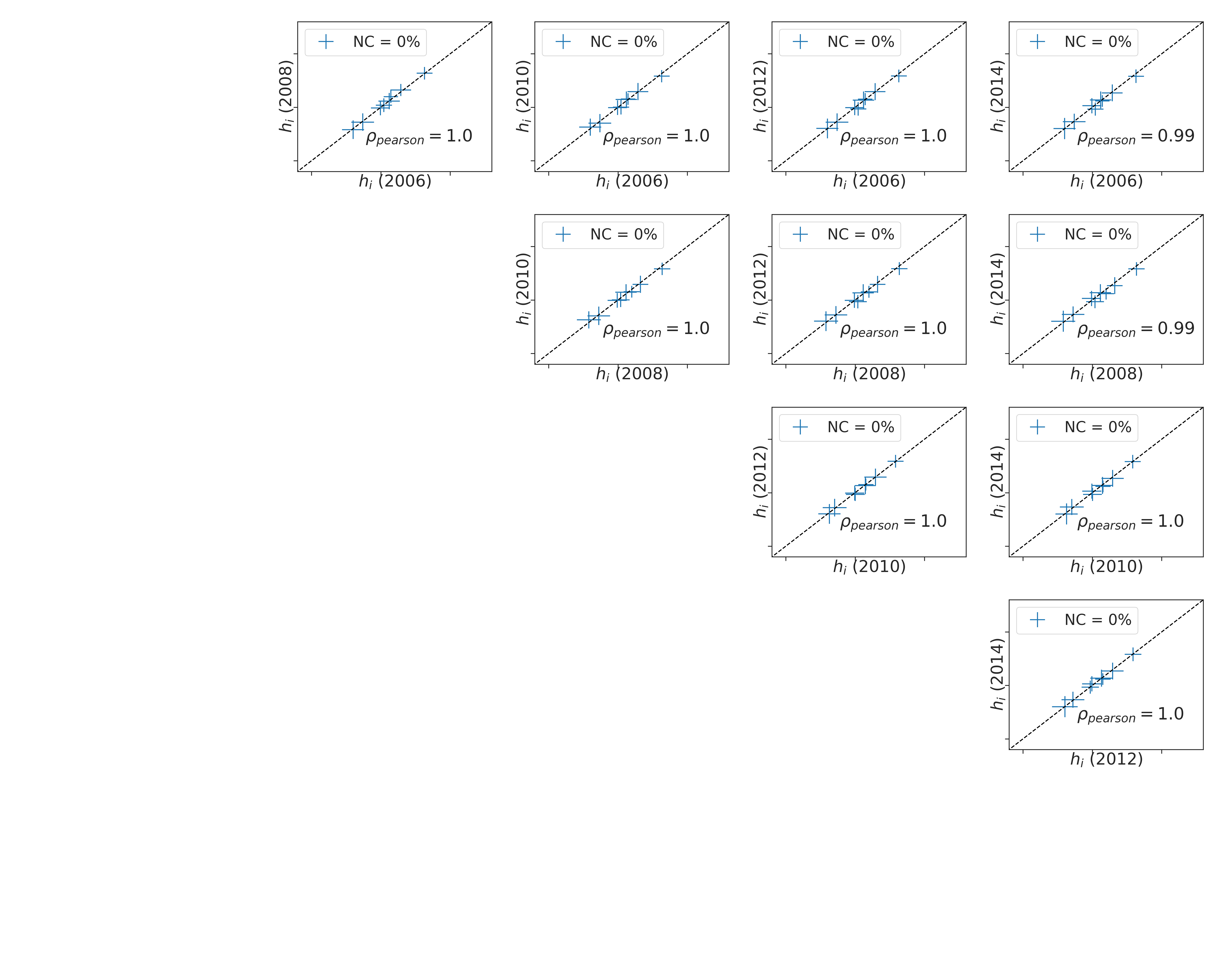}
	\caption{Comparisons between the components of $J^{(1)}$ in different years. Errors of each parameter are reported in the plot. Dotted line represents the identity relation. The percentage of component with a $p$-value from $t$-test below $0.05$ is shown in the legend.}
	\label{figsi:J1_comparisons}
\end{figure}
\FloatBarrier
\section{Maximum Likelihood Estimation of the $dt$ parameter}
Assuming that our system can be described by a Langevin equation in the form of
\begin{equation}
\frac{dx}{dt}(t)= -\nabla H (x) + \eta(t),
\label{eqsi:langevin}
\end{equation}
it is easy to derive a discrete version of this equation, capable of coping with the discrete nature of the data we have. Supposing to have a small shift in time $dt$ and calling $t' = t + dt$ we have
\begin{equation}
x(t')= x(t) -\nabla H (x(t)) dt/2 + \eta \sqrt{dt}.
\label{eqsi:langevin_discrete}
\end{equation}
We assumed in the main text that the each component of the noise $\eta_i$ is distributed according to a Laplace distribution with variance $1$ and that different components are uncorrelated. Hence, each component of the vector $(x(t') - x(t) + \nabla H(x(t))dt)/\sqrt{dt}$ will be a Laplace-distributed variable. This simple fact allows to compute the transition probability from $x(t)$ to $x(t')$, that will be in the form
\begin{equation}
\begin{split}
&\P_{dt}(x(t') | x(t) )= \\& \prod_{j=1}^N \sqrt{\frac{1}{2dt}} \exp \left (  -\frac{ \sqrt{2} | x_j(t+dt) - x_j(t) - (\partial H / \partial x_j)(\vec{x}(t)) dt / 2|}{\sqrt{dt}} \right).
\end{split}
\label{eqsi:small_gauss_jump}
\end{equation}
At this point we would like to match the intrinsic time $t$ of the model, with the real time of the data. To do so, we need to understand which $dt$ corresponds to a time frame of one year. We can use Maximum Likelihood to fix this value, trying to maximize the Log-likeihood obtained by applying (\ref{eqsi:small_gauss_jump}). In other words, we look for the value of $dt$ maximizing the probability of observing the transitions we have in the data. Such log-likelihood can be written as
\begin{equation}
\mathcal{L}(dt) = \frac{1}{4 N_C} \sum_\alpha \sum_{y=2006}^{2015} \P_{dt}(x^\alpha(t_{y+1})  | x^\alpha(t_y) ),
\label{eqsi:small_gauss_jump_bis}
\end{equation}
where $x^\alpha(t_y)$ is the vector of indicators of the city $\alpha$ in the year $y$ (the notation $t_y$ indicates the intrinsic time corresponding to the year $y$). Fig.~ show log-likelihood as a function of $dt$. The maximum observed value of $\mathcal{L}$ has been found for $dt_{\textit{max}} = 0.014$.
\begin{figure}    
	\centering
	\includegraphics[width=0.5\linewidth]{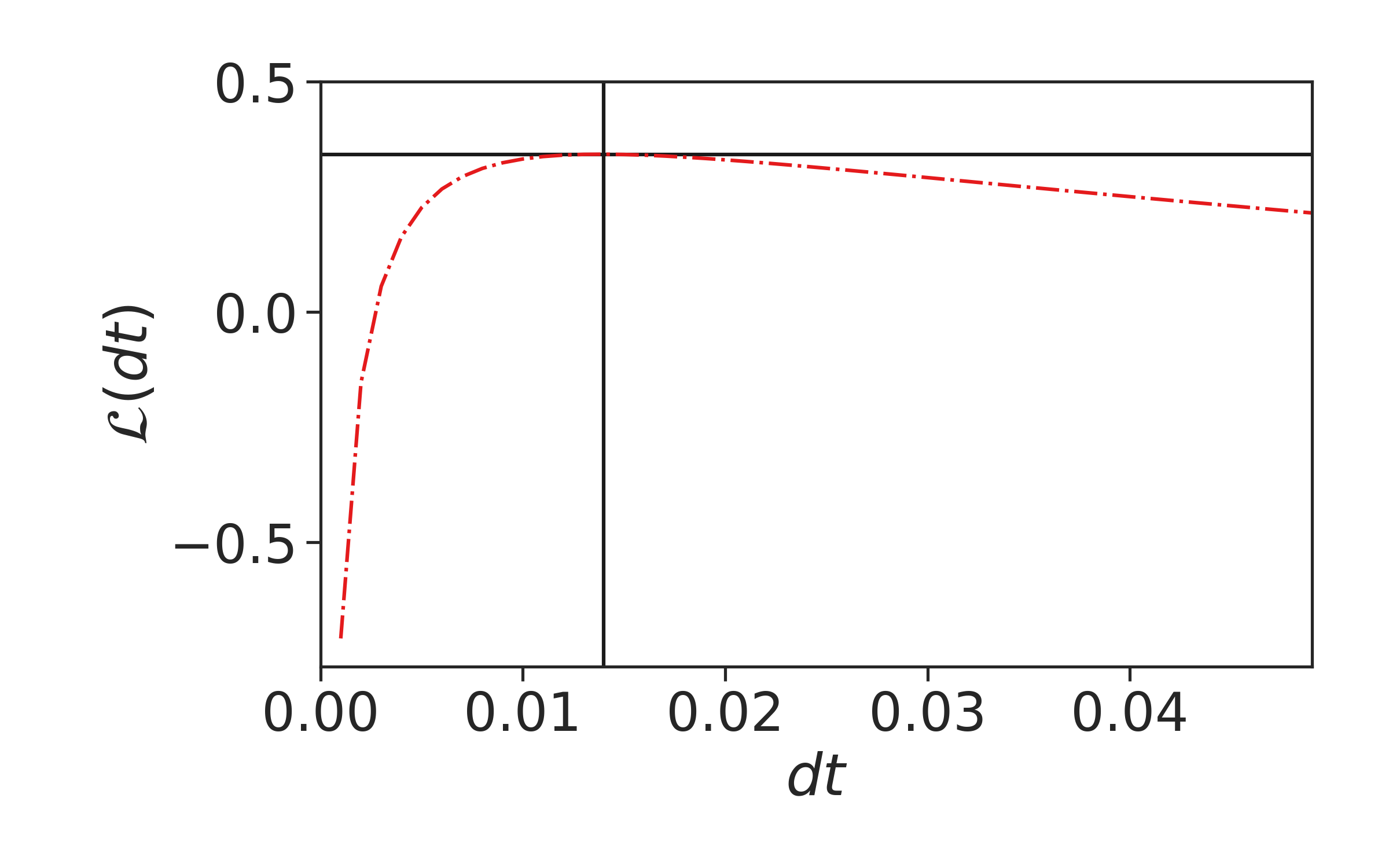}
	\caption{Log-likelihood in equation (\ref{eqsi:small_gauss_jump_bis}) as a function of $dt$. The maximum observed value of $\mathcal{L}$ is highlighted in the plot and has been found at $dt_{\textit{max}} = 0.014$.}
	\label{fig:ML_dt}
\end{figure}
\FloatBarrier
\section{Comparison with Causal Inference}
The static model inferred in the main text is capable of predicting the evolution of a city if we use its corresponding Langevin equation to define a dynamics. In this case, we use the temporal information in the data only to infer the parameter $dt$ used to make the Langevin equation discrete. Another approach we can use to define dynamic models is to use temporal correlations explicitly according to the Maximum Caliber principle\cite{presse2013principles}. First, we need to define time-dependent observables, i.e. observables depending on variable at different times.
For sake of simplicity, we will focus on correlations of order $2$ defined as, 
\begin{equation}
C^{(2)}_{i,j}(\delta)= \langle x_{i}(y+\delta)x_{j}(y) \rangle_{\textit{data}},
\label{sieq:n_corr_time}
\end{equation}
where now the average is taken over all the communes in the data-set and all the years. As observables for the definition of the model we choose $C^{(1)}_{i}$, $C^{(2)}_{i,j}(\delta=1)$. In this way, we are modelingng explicitly the average of the sample and the correlations between the indicators in consecutive years. The model corresponding to this set of observables has a transition probability defined by
\begin{equation}
\begin{split}
& \P(x(y+1)|x(y)) \propto\\   
& \exp  \left  ( -\sum_i \frac{x_i(y+1)^2}{2}  - \sum_{ij}  B_{ij} x_i(y+1)x_j(y) + \sum_i c_i x_i(y+1) \right ).
\end{split}
\label{sieq:causal_prob}
\end{equation}
This model corresponds to a linear model defined as 
\begin{equation}
x(y+1) = -B x(y) + h + \eta
\label{sieq:causal_dyn}
\end{equation}
where $\eta$ is a normally distributed random variable with mean equal to $0$ and variance equal to $1$. Being equation (\ref{sieq:causal_dyn}) corresponding to a linear model, its parameters can be inferred by a standard linear regression. \newline
\FloatBarrier
\end{document}